\documentclass[a4paper,11pt]{article}
\pdfoutput=1 
\usepackage{jheppub}
\usepackage{mathrsfs}
\usepackage{amsmath}
\usepackage{amssymb}
\usepackage{subcaption}
\usepackage{amsfonts}
\usepackage[toc,page]{appendix}
\usepackage[T1]{fontenc} % if needed
\usepackage{breqn}
\usepackage{caption}
\usepackage{subcaption}
\usepackage{comment}
\excludecomment{hidden}
\usepackage{caption}

\title{\boldmath Restricted Phase Space Thermodynamics of Dyonic AdS Black Holes: Comparative Analysis Using  Different  Entropy Models}

\author{ Abhishek Baruah, }
\author{Prabwal Phukon}
\affiliation{Department Of Physics,\\Dibrugarh University, Dibrugarh,786004, Assam, India}
\affiliation{Theoretical Physics Division, Centre for Atmospheric Studies, Dibrugarh University, Dibrugarh, 786004, Assam, India}

\emailAdd{rs{\_}abhishekbaruah@dibru.ac.in}
\emailAdd{prabwal@dibru.ac.in}

\abstract{Using the Restricted Phase Space (RPST) formalism, we perform a comparative study of 4D dyonic AdS black hole thermodynamics in Gibbs-Boltzmann statistics and R\'enyi statistics. In RPST formalism, instead of pressure and volume, one considers central charge $C$ and chemical potential $\mu$ as thermodynamic variables. Inclusion of the magnetic charge $\tilde{Q}_m$ gives rise to a richer phase structure of the study of thermodynamics by adding a non-equilibrium transition from an unstable small black hole to a stable black hole on top of the Van der Waals transition in the $T-S$ processes and a Hawking-Page transition in the $F-T$ plots. We study an extra mixed ensemble ($\tilde{\Phi}_e,\tilde{Q}_m)$ due to the inclusion of $\tilde{Q}_m$ where we see Van der Waals phase transition and whose plots change as the entropy model changes though the style of transition remains the same. We observe an interesting phenomenon where changing the R\'enyi parameter $\lambda$, the $T-S$ process changes the same way as when varying the central charge $C$ underlining some similarity that is not seen in the Bekenstein Hawking entropy model. We observe a similarity between the plots when both charges are turned off relating to the Schwarzschild black hole and the grand-canonical ensemble. One can observe that as the entropy models are changed, the homogeneity is not lost where the mass as a function of extensive variables is of order one and the rest zero. Finally, we see a similarity in the $\mu-C$ process across the entropy models signally some universality across entropy models as well as different types of black holes studied before. }

\begin{document} 
\maketitle
\flushbottom

\section{Introduction}
\label{sec:one}
Black holes are cosmic enigmatic objects or entities where the gravity is so strong that it warps both time and space beyond our comprehension. It continues to offer us profound insights into the vivid nature of spacetime, quantum mechanics, and general relativity by continuing to challenge us in its understanding. Since the inception of the study of black hole mechanics, certain laws were pioneered by Bardeen Carter, Bekenstien, and Hawking \cite{a,aa1,b,bb1} which led to the study and investigation of black hole thermodynamics by considering the Hawking temperature $T$ and Bekenstein-Hawking entropy S and currently it is an active area of research \cite{0a,0b,0c,0d,0e}. The study of black holes' phase transition is quite fascinating and can be traced to originated from the study of Anti-de Sitter black holes by Hawking and Page where they discovered a phase transition between the pure thermal AdS space and Schwarzschild AdS black hole \cite{d}. There are now even more ways to study the phase transitions \cite{1a,1b,1c} via the thermodynamic geometry \cite{2a,2b,2c,2d,2e,2f,2g,2h,2i,2j,2k} and the topology study of the black holes \cite{3a,3b,3c,3d,3e,3f,3g,3h,3i,3j,3k,3l,3m,3n}.
The AdS black hole also has a one-to-one correspondence with the Conformal Field Theory CFT which is typically known as the AdS/ CFT correspondence \cite{c}. In this light the phase transition of Hawking and Page in the bulk side associates with the confinement/de-confinement phases of the quark-gluon plasma in the boundary side \cite{dd1}. It is also seen that a liquid/gas transition of the Van der Waals fluid looks like a change of phase of order one for the small/large black holes having charge, rotating, or both.\\
The extended Phase Space Thermodynamics also known as the EPST \cite{i,x,y,yy,yy1,yy2,yy3,yy4,j1,p} where the positive pressure $P$ is related to the negative cosmological constant $\Lambda (<0)$ where $P=-\Lambda/8 \pi G$ and $\Lambda=-(d-1)(d-2)/2l^2$ where $l$ is the curvature radius introduces a new pair of state variables $(P,V)$ and be able to be considered as heat engines \cite{b1,bc1}. With this idea in the picture, black holes are portrayed as such systems that have a dual correspondence to other studies like quantum chromodynamics, and condensed matter physics apart from the Conformal Field Theory  \cite{f,g,h,j,o,m,n,k,l}.\\
 Research in the thermodynamics of black holes using the extended phase space thermodynamics is said to be black hole chemistry \cite{j1,p}. Visser's formalism \cite{u} is a work using the EPST formalism which gets its contribution from the AdS/CFT correspondence by Maldacena \cite{c}. Its contribution is the addition of the CFT's central charge $C$ and its conjugate chemical potential $\mu$ in the thermodynamics first law  \cite{v,vv}. Both the pressure and the black hole's pressure are converted to the CFT's volume and pressure using the AdS/CFT dictionary. In the Visser formalism of CFT's volume and pressure $(p,\mathcal{V})$, the AdS radius shows a more significant role. With all the points taken, we can now write the first law for the charged rotating black hole as
 \begin{equation}
 dE=TdS-p \mathcal{V}+\tilde{\Phi}\tilde{Q}+\Omega J+\mu C
 \end{equation}
 where we also get the Smarr relation as $E=TS+\tilde{\Phi}\tilde{Q}+\Omega J+\mu C$ where $\tilde{\Phi}$ and $\tilde{Q}$ are both the rescaled potential and charge. There is a full description of the bulk thermodynamics and the CFT boundary thermodynamics in the framework of Visser. But a problem comes when replacing the CFT's $(p, \mathcal{V})$ with the thermodynamic $(P,V)$ and can be avoided by fixing the central charge $C$ and expressing the black hole mass as the enthalpy.\\
 There seems to be an absence of the homogeneity relation of the intensive variables with the internal mass in the EPST  and the proper explanation of the volume term both in the first law and the Smarr relation as well as the Gibbs-Duhem relation is not seen as well as there is a missing $p \mathcal{V}$ term in the Smarr relation. On top of it, Visser's initial formalism has an issue called the $\textit{ensemble of theories}$. This can be bypassed if one assumes a strict version of Visser's formalism which is called the $\textit{restricted phase space thermodynamics}$ \cite{rps}.
 The RPST formalism has been studied for various black holes \cite{rps1,rps2,rps3,rps4,rps5}. The RPST formalism provides us with various insights and rich phase structures because it has multiples degrees of freedom even though it doesn't have the $PdV$ term in the first law.\\
 In this paper, we have used the RPST formalism on a dyonic black hole where it has two fields: one is the graviton (metric) and the other is the $U(1)$ gauge field \cite{l2}. A dyonic black hole metric satisfies two boundary conditions and one of them leads to the deformation of the boundary theory this deformed $(2+1)$ dimension theory has led to many studies \cite{aa,ab,ac,ac1,ac2,ad,z,a1,d1,e1,f1,c1}. The addition of a magnetic monopole adds to the enrichment of the phase structures and broadens and shines more light on the thermodynamics of black holes. Adding more variables will add to more degrees of freedom making the phase transition more vibrant and interesting.\\
 Over the years on the path to developing quantum mechanics, quantum gravity, and non-extensive thermodynamics, it is seen that entropy is not quite fundamental and shares its dependency on the physical system and changes by varying the physical theories. The Bekenstein-Hawking entropy from the start was proportional to the horizon area rather than the volume. The systems entropy in our classical thermodynamics depends on the Mass and volume and so it is an additive and extensive quantity. The idea why the black hole entropy is non-extensive is still under study \cite{en1} and therefore alternative studies have been done having different constructs depending on the non-extensive criteria \cite{en1,en2,en3,en4,en5,en6,en7,en8,en9,en10,en11,en12,en13,en14,en15,en16,en17,en18}. As the first law of black hole thermodynamics depends on the entropy, temperature, energy, and other quantities, changing the entropy will change the thermodynamics of the system \cite{en19}. Some alternate constructs are replacing the Bekenstein entropy like the R\'{e}nyi \cite{en20}, Tsallis \cite{en21}, Barrow \cite{en22}, Sharma-Mittal \cite{en23} and the recent Kaniadakis proposals \cite{en24,en25}.\\
 Our motivation in this paper is to study the dyonic AdS black holes in the Restricted Phase Space Thermodynamics which due to the inclusion of a magnetic monopole extends or opens new doors for more ensemble interplay thereby making a bigger platform for more phase structures not seen before and with this by studying in different entropy constructs can show us any similarity with the entropy constructs or any uniqueness with a certain model. We take here the usual Bekenstein-Hawking entropy where many studies have been done using the RPST for various black hole systems but not seen for the dyonic black hole as well we added a study with the R\'enyi entropy as well to give us a better visualization and understanding of the relation of different non-extensive entropy constructs within the RPST formalism.\\
 The format of the paper is as follows. In Section \ref{sec:two}, we review the thermodynamics of dyonic black holes using the Restricted Phase Space formalism. In Section \ref{sec:three}, we discuss the equation of states and states of homogeneity using the Bekenstein-Hawking entropy. In Section \ref{sec:four} We discuss the equation of states and homogeneity using the R\'{e}nyi entropy construct. In Section \ref{sec:five} we conclude the paper and write the overall conclusion pointwise. Section \ref{appendix} is the Appendix.

\section{Thermodynamics of Dyonic black hole in the Restricted Phase Space}
\label{sec:two}
In the presence of the cosmological term, we can vary the Reissner-Nordstr$\ddot{o}$m action and arrive at a solution that solves the equation of motion.\\
In the presence of the negative cosmological constant, we take the Reissner-Nordstr$\ddot{o}$m action in D=4
\begin{equation}
I=\frac{1}{16 \pi G_4}\int d^4x\sqrt{g}\left(-R+F^2-\frac{6}{b^2}\right)
\end{equation}
Then, we write the equation of motion as
\begin{equation}
R_{\mu \nu}-\frac{1}{2}g_{\mu \nu}R-\frac{3}{b^2}g{\mu \nu}=2(F_{\mu \lambda} F_\nu^\lambda-\frac{1}{4}g_{\mu \nu}F_{\alpha \beta}F^{\alpha \beta})
\end{equation}
\begin{equation}
\Delta_\mu F^{\mu \nu}=0
\end{equation}
A static symmetric solution is given as
\begin{equation}
A=\left(\frac{-Q_e}{r}+\frac{Q_e}{r_+}\right)dt+(Q_m\cos\theta)d\phi
\end{equation}
and the metric is given as
\begin{equation}
ds^2=-f(r)dt^2+\frac{1}{f(r)}dr^2+r^2d\theta^2+r^2\sin^2\theta d\phi^2
\end{equation}
where 
\begin{equation}
f(r)=\left(1+\frac{r^2}{b^2}-\frac{2GM}{r}+
\frac{G(Q_e^2+Q_m^2)}{r^2}\right)
\end{equation}
where  $Q_e$ is the electric charge, $Q_m$ is the magnetic charge $M$ is the mass of the black hole and the horizon is $r_+$. 
By setting $f(r_+)=0$ we can get the mass as
\begin{equation}
\label{eq:mass1}
M=\frac{G l^2 Q_e^2+G l^2 Q_m^2+l^2 r^2+r^4}{2 G l^2 r}
\end{equation} 
In the Restricted phase space thermodynamics of dyonic black holes, the four pairs of variables: $(S,T)$, $(\tilde{Q}_e,\tilde{\Phi}_e)$, $(\tilde{Q}_m, \tilde{\Phi}_m)$, and $(C,\mu)$ that are conjugate forms the macro states. The entropy $S$ and the conjugate  temperature $T$ is given as
\begin{equation}
S=\frac{A}{4G}=\frac{\pi r_+^2}{G}, \quad T=\frac{r^2 \left(l^2+3 r_+^2\right)-G l^2 \left(Q_e^2+Q_m^2\right)}{4 \pi  l^2 r_+^3}
\end{equation}
Now we write the rescaled electric charge $\tilde{Q}_e$ and the coloumb potential $\tilde{\Phi}_e$ are given as
\begin{equation}
\label{eq:rcharge1}
\tilde{Q}_e=\frac{Q_e l}{\sqrt{G}}, \quad \tilde{\Phi}_e=\frac{\Phi_e \sqrt{G}}{l}=\frac{Q_e \sqrt{G}}{r_+ l}
\end{equation}
and the rescaled magnetic charge $\tilde{Q}_m$ and magnetic potential $\tilde{\Phi}_m$ are given as
\begin{equation}
\label{eq:rcharge2}
\tilde{Q}_m=\frac{Q_m l}{\sqrt{G}}, \quad \tilde{\Phi}_m=\frac{\Phi_m \sqrt{G}}{l}=\frac{Q_m \sqrt{G}}{r_+ l}
\end{equation}
Finally, the central charge $C$ and its conjugate chemical potential $\mu$ is given as
\begin{equation}
\label{eq:mu}
C=\frac{l^2}{G}, \quad \mu=\frac{M-TS-\tilde{\Phi}_e\tilde{Q}_e-\tilde{\Phi}_m\tilde{Q}_m}{C}
\end{equation}
Be calculations, one could verify that
\begin{equation}
\label{eq:first}
dM=TdS+\tilde{\Phi}_e\tilde{Q}_e+\tilde{\Phi}_m\tilde{Q}_m+\mu C
\end{equation}
Rearranging equation \eqref{eq:mu}, we get
\begin{equation}
\label{eq:smarr}
M=TS+\tilde{\Phi}_e\tilde{Q}_e+\tilde{\Phi}_m \tilde{Q}_m+\mu C
\end{equation}
Using \eqref{eq:first}, \eqref{eq:smarr} we get the Gibbs-Duhem relation
\begin{equation}
d \mu=-\tilde{\mathcal{Q}}d\tilde{\Phi}-\mathcal{S}dT, \quad \tilde{\mathcal{Q}}=\tilde{Q}/C, \quad \mathcal{S}=S/C
\end{equation}
where $\tilde{\mathcal{Q}}$
is the $e$-charge per unit $C$-charge and and $\mathcal{S}$ is the entropy per unit $C$-charge.\\
It is important to note that though the variables central charge C and the $\mu$ are borrowed from the dual CFT, they can be interpreted as the effective number of microscopic degrees of freedom $N_{bulk}$ and as the chemical potential of the black hole in the bulk. We propose that $\mu_{CFT}=\mu_{bulk}$, $C=N_{bulk}$ can seem to serve as a bridge within the holographic dictionary. For clarity in the notation, we shall retain the original notation $(\mu,C)$ instead of replacing it with $(\mu_{bulk}, N_{bulk})$, while emphasizing that our entire focus is on the thermodynamics of the bulk rather than its CFT. 
\section{Equation of states and homogeneity using the Bekenstein-Hawking Entropy}
\label{sec:three}
The Bekenstein-Hawking entropy is given as
\begin{equation}
S=\frac{A}{4G}=\frac{\pi r_+^2}{G}
\end{equation}
Solving this entropy we get the value of $r$ as
\begin{equation}
r= \frac{\sqrt{G} \sqrt{S}}{\sqrt{\pi }}
\end{equation}
Putting this value of $r$ in  mass equation \eqref{eq:mass1} we get
\begin{equation}
M=\frac{G S^2+\pi  l^2 \left(\pi  \left(Q_e^2+Q_m^2\right)+S\right)}{2 \pi ^{3/2} \sqrt{G} l^2 \sqrt{S}}
\end{equation}
Then we substitute the electric charge and magnetic charge with the rescaled electric charge and rescaled magnetic charge given in \eqref{eq:rcharge1} and \eqref{eq:rcharge2} and the mass equation is written as:-
\begin{equation}
M=\frac{G \left(\pi ^2 \left(\tilde{Q}_e^2+\tilde{Q}_m^2\right)+S^2\right)+\pi  l^2 S}{2 \pi ^{3/2} \sqrt{G} l^2 \sqrt{S}}
\end{equation}
Then putting the value of Newton's constant $G_N$ in terms of the central charge $C$ and the AdS length $l$ as in \eqref{eq:mu}, we get
\begin{equation}
\label{eq:cftmass}
M=\frac{ \left(4 \pi  C S+\pi ^2 \left(\tilde{Q}_e^2+\tilde{Q}_m^2\right)+S^2\right)}{4 \pi ^{3/2} l \sqrt{C S}}
\end{equation}
From the first law given in equation \eqref{eq:first}, we can derive all the equations of states by simply partially derivative of $M$ and we get
\begin{equation}
\label{eq:cftall}
\begin{split}
&T=\frac{ \left(4 \pi  C S-\pi ^2 \left(\tilde{Q}_e^2+\tilde{Q}_m^2\right)+3 S^2\right)}{8 \pi ^{3/2} l C^{1/2} S^{3/2}}\\
&\tilde{\Phi}_e=\frac{\sqrt{\pi } \tilde{Q}e }{2 l \sqrt{CS}}\\
&\tilde{\Phi}_m=\frac{\sqrt{\pi } \tilde{Q}_m }{2 l \sqrt{CS}}\\
&\mu=\frac{4 \pi  C S-\pi ^2 \left(\tilde{Q}_e^2+\tilde{Q}_m^2\right)-S^2}{8 \pi ^{3/2} C^{3/2} l \sqrt{S} }
\end{split}
\end{equation}
We can check whether the mass equation is of first order and the rest is of zeroth order. All this is achieved by rescaling $S\rightarrow \lambda S$, $\tilde{Q}_e \rightarrow\lambda\tilde{Q}_e$, $\tilde{Q}_m\rightarrow\lambda \tilde{Q}_e$, $C \rightarrow \lambda C$ firstly in equation \eqref{eq:cftmass} and then in \eqref{eq:cftall} we get
\begin{equation}
M=\frac{\lambda \left(4 \pi  C S+\pi ^2 \left(\tilde{Q}_e^2+\tilde{Q}_m^2\right)+S^2\right)}{4 \pi ^{3/2} C \sqrt{\lambda S} \sqrt{\frac{l^2}{\lambda C}}} \implies M=\lambda M
\end{equation}
where the Mass $M$ is also rescaled as $\lambda M$ which is indicating that there is a first order homogenity of $M$ and
\begin{equation}
\begin{split}
&T=\frac{\lambda \left(4 \pi  C S-\pi ^2 \left(\tilde{Q}_e^2+\tilde{Q}_m^2\right)+3 S^2\right)}{8 \pi ^{3/2} C (\lambda S)^{3/2} \sqrt{\frac{l^2}{\lambda C}}} \implies T=\lambda^0 T\\
&\tilde{\Phi}_e=\frac{\sqrt{\pi } \lambda  \tilde{Q}_e \sqrt{\frac{1}{C \lambda }}}{2 l \sqrt{\lambda  S}} \implies \tilde{\Phi}_e=\lambda^0 \tilde{\Phi}_e\\
&\tilde{\Phi}_m=\frac{\sqrt{\pi } \lambda  \tilde{Q}_m \sqrt{\frac{1}{C \lambda }}}{2 l \sqrt{\lambda  S}} \implies \tilde{\Phi}_m=\lambda^0 \tilde{\Phi}_m\\
&\mu=\frac{4 \pi  C S-\pi ^2 \left(\tilde{Q}_e^2+\tilde{Q}_m^2\right)-S^2}{8 \pi ^{3/2} C^2 \sqrt{\lambda  S} \sqrt{\frac{l^2}{C \lambda }}} \implies \mu=\lambda^0 \mu
\end{split}
\end{equation}
We see that the remaining variables do not get rescaled like  that $M$, which  shows a zero-order homogeneity of $T$, $\tilde{\Phi}_e$, $\tilde{\Phi}_m$ and $\mu$.
\subsection{Thermodynamic processes and phase transitions}
\begin{figure}[h]
    \centering
    \begin{subfigure}[b]{0.4\textwidth}
        \includegraphics[scale=0.7]{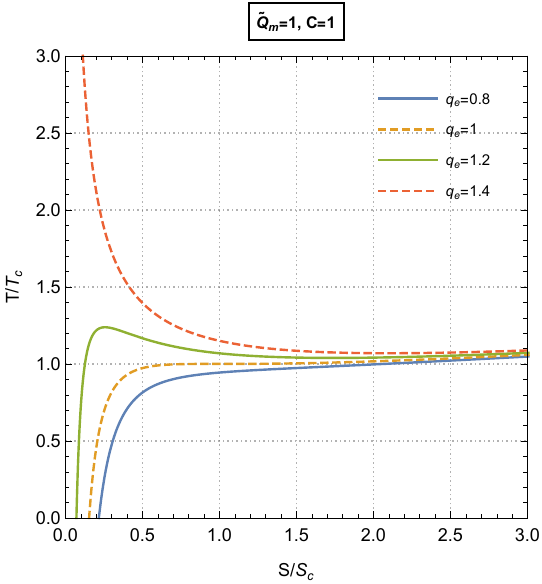}
    \end{subfigure}
    \hfill
    \begin{subfigure}[b]{0.4\textwidth}
        \includegraphics[scale=0.7]{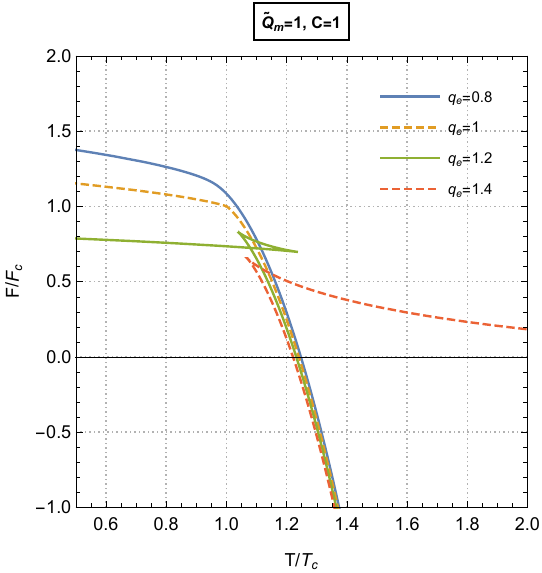}
    \end{subfigure}
    \caption{$T-S$ and $F-T$ plots for various values of $q_e$}
    \label{fig:one}
\end{figure}

\begin{figure}[h]
\centering
\includegraphics[scale=0.8]{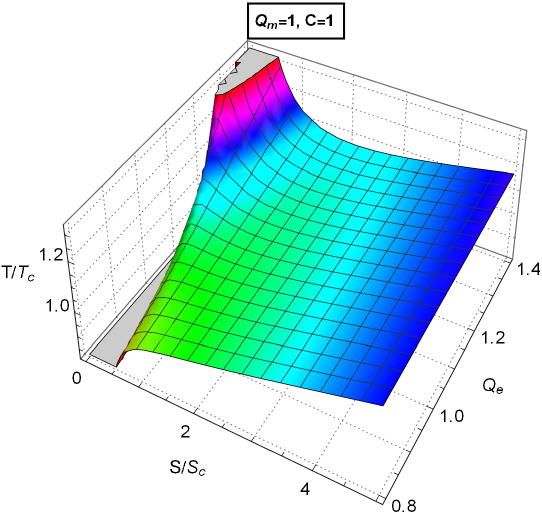}
\caption{3-d plot for $T-S-\tilde{Q}_e$ }
\label{fig:3d11}
\end{figure}

\begin{figure}
\centering
\includegraphics[scale=0.8]{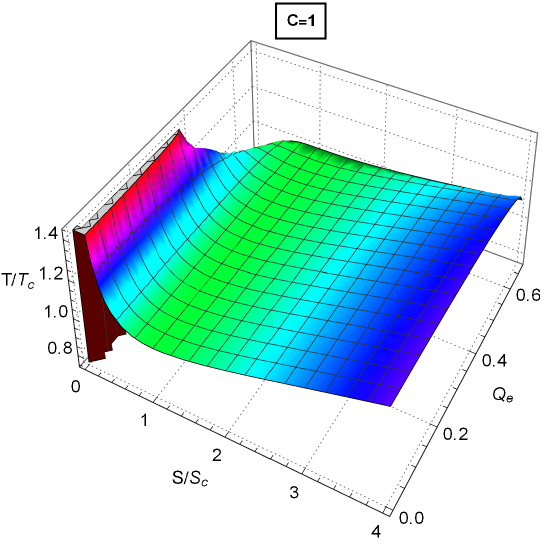}
\caption{3-d plot for $T-S-\tilde{Q}_e$}
\label{fig:3d2}
\end{figure}

\begin{figure}[htp]
    \centering
    \begin{subfigure}[b]{0.4\textwidth}
        \includegraphics[scale=0.7]{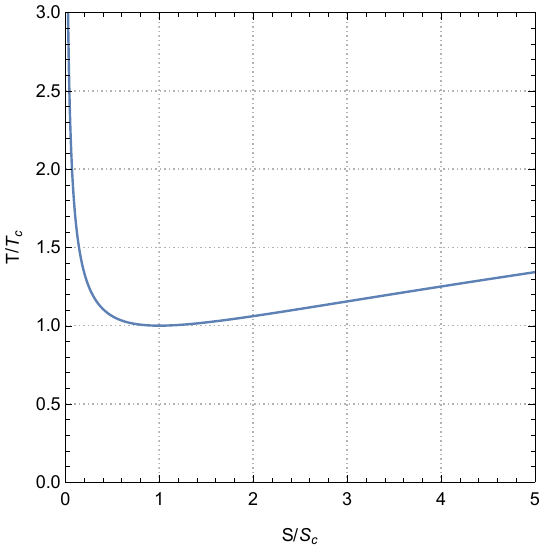}
    \end{subfigure}
    \hfill
    \begin{subfigure}[b]{0.4\textwidth}
        \includegraphics[scale=0.7]{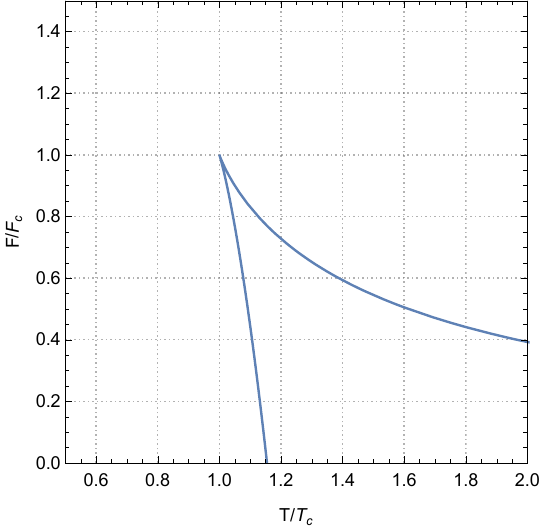}
    \end{subfigure}
    \caption{$T-S$ and $F-T$ plots for $\tilde{Q}_e=\tilde{Q}_m=0$}
    \label{fig:three}
\end{figure}

In this section, we study the various thermodynamic processes for the dyonic AdS black hole in the RPST formalism by using the Bekenstein-Hawking entropy.
We can see that the state of the dyonic AdS black hole can be viewed by the four pairs of thermodynamic variables namely $(T,S)$, $(\tilde{\Phi}_e,\tilde{Q}_e)$, $(\tilde{\Phi}_m,\tilde{Q}_m)$, $(\mu,C)$ and if anyone parameter changes slightly, we can say that a macro process has occurred.\\
First we try the check the $T-S$ plots for the fixed $\tilde{Q}_e$, $\tilde{Q}_m,$ $C$. When $T=0$ then we can calculate and upper bound for electric charge $\tilde{Q}_e$ for numerous values of $\tilde{Q}_m$, $S$ and $C$.
\begin{equation}
\tilde{Q}_e=\tilde{Q}_e^{\text{max}}=\frac{ \sqrt{4 \pi  C S-\pi ^2 \tilde{Q}_m^2+3 S^2}}{\pi }
\end{equation}
Now for the $T-S$ plots, we can find the point of inflection by using certain equations given as:-
\begin{equation}
\label{eq:differential}
\left(\frac{\partial T}{\partial S}\right)_{\tilde{Q}_e, \tilde{Q}_m,C}=0, \quad \left(\frac{\partial^2 T}{\partial S^2}\right)_{\tilde{Q}_e, \tilde{Q}_m,C}=0
\end{equation}
After using the above equation, we get a point where the $\tilde{Q}_e$ takes the crucial value as:-
\begin{equation}
\label{eq:critqq}
\tilde{Q}_e=\tilde{Q}_e^{c}\equiv \frac{1}{3} \sqrt{4 C^2-9 \tilde{Q}_m^2}
\end{equation}
where we can put various values of the central charge $C$ and $\tilde{Q}_m$ to get a particular crucial point. There the critical values for $S$ and $T$ can be given as:-
\begin{equation}
\label{eq:crits}
S_c=\frac{2 \pi C}{3}, \quad T_c=\frac{\sqrt{2}}{\sqrt{3} l \pi}
\end{equation}
We introduce some relational variables:-
\begin{equation}
t=\frac{T}{T_c},\quad s=\frac{S}{S_c}, \quad q_e=\frac{\tilde{Q}_e}{\tilde{Q}_e^c}
\end{equation}
Using this, the temperature equation at \eqref{eq:cftall} can be written as:-
\begin{equation}
\label{eq:temp1}
t=\frac{4 \left(-q_e^2+3 s^2+6 s\right)+9 \left(\tilde{Q}_m^2-1\right) \tilde{Q}_m^2}{32 s^{3/2}}
\end{equation}
It is interesting to see that the above equation of state is devoid of the dependency of the central charge $C$. This type of thermodynamic study is called the law of corresponding states, related to the study of ordinary matter.\\
The Helmholtz free energy is given by using \eqref{eq:cftmass}, \eqref{eq:cftall}, we get:-
\begin{equation}
F(T,\tilde{Q}_e, \tilde{Q}_m,C)=M(S,\tilde{Q}_e, \tilde{Q}_m,C)-TS
\end{equation}
By using the limiting values in \eqref{eq:critqq} and \eqref{eq:crits}, we get the critical value of $F$ and also introduce a relative variable for the free energy as:-
\begin{equation}
F_c=\frac{2 \sqrt{2}}{\sqrt{3} l}, \quad f=\frac{F}{F_c}
\end{equation}
Hence, the value of the parameter $f$ is
\begin{equation}
\label{eq:free1}
f=\frac{-4 \left(-3 q_e^2+s^2-6 s\right)-27 \left(q_e^2-1\right) \tilde{Q}_m^2}{32 \sqrt{s}}
\end{equation}
We can notice that the free energy $f$ parameter is also devoid of the central charge $C$ expression.\\
We plot the $T-S$ and $F-T$ relating to the iso-$e$-charge process plot using \eqref{eq:temp1}, \eqref{eq:free1} shown in Figure \ref{fig:one}.
In Figure \ref{fig:one}, we see for the points below the critical points $q_e$, we see that the plot is monotonically increasing, it is similar to the $F-T$ graph too. Now at any supercritical value of $q_e$, we see the curve becomes non-monotonic and in the $F-T$ graph, we see a swallowtail structure where it is cutting at the bottom of the plot at some temperature where $T>T_c$. This kind of appearance shows a Van der Waals type phase transition of order one present in the iso-$e$-charge processes where $0<\tilde{Q}_e<\tilde{Q}_e^c$
This above process is not related to any volume work because we haven't linked that black hole with the idea of volume. At $\tilde{Q}_e=\tilde{Q}_e^c$ we get the change of phase of order two.\\
We observe that we see a change of phase of order one above the crucial value of $\tilde{Q}_e$, which was also seen in the EPST formalism.
Also, in Fig \ref{fig:one}, we see for the iso-$e$-charge $T-S$ plot, that plots for some values of $\tilde{Q}_e$ decreases as the temperature $T$ is decreased and cuts the $S$ axis as some point which relates to a black hole remnant. But if we observe the red graph, as the temperature decreases, the entropy increases until it reaches a stable non-vanishing value marking a non-equilibrium transition from an unstable small black hole to a stable large black hole. Correspondingly, in the $F-T$ plot for the same value of $\tilde{Q}_e$, we get the Hawking-Page type of phase transition as seen for the red plot in Figure \ref{fig:one}. We can also see the 3-d plot for $T-S-\tilde{Q}_e$ keeping $C$ and $\tilde{Q}_m$ fixed in Figure \ref{fig:3d11}.\\
Now we can also see by plotting the magnetic charge $\tilde{Q}_m=0$. But setting that condition, the $T-S$ and the $F-T$ plot becomes a bit different, namely the non-equilibrium transition and Hawking-Page type phase transition disappears in the respective plots.
By using \eqref{eq:cftall} and by setting $\tilde{Q}_m=0$, we can calculate the temperature as:- 
\begin{equation}
T=\frac{ \left(4 \pi  C S-\pi ^2 \tilde{Q}_e^2+3 S^2\right)}{8 \pi ^{3/2} l C^{1/2} S^{3/2}}
\end{equation}
Now by using the equation \eqref{eq:differential}, we get the critical values of entropy $S$ and the electric charge $\tilde{Q}_e$ as:-
\begin{equation}
S_c=\frac{2 C \pi}{3}, \quad \tilde{Q}_e^c=\frac{2C}{3}
\end{equation}
Using these critical points, the temperature and the free energy can be rescaled to the equation
\begin{equation}
\label{eq:TF}
t=\frac{\left(-q_e^2+3 s^2+6 s\right)}{8 s^{3/2}}, \quad f= -\frac{ \left(-3 q_e^2+s^2-6 s\right)}{8 \sqrt{ s}}
\end{equation}
where $t=T/T_c$, $f=F/F_c$ are the temperature and free energy parameter where $T_c=\sqrt{\frac{2}{3}}\frac{1}{l\pi}$ and $F_c=\frac{2 \sqrt{2} C}{3\sqrt{3} \sqrt{l}}$.
\begin{figure}[htp]
    \centering
    \begin{subfigure}[b]{0.4\textwidth}
        \includegraphics[scale=0.7]{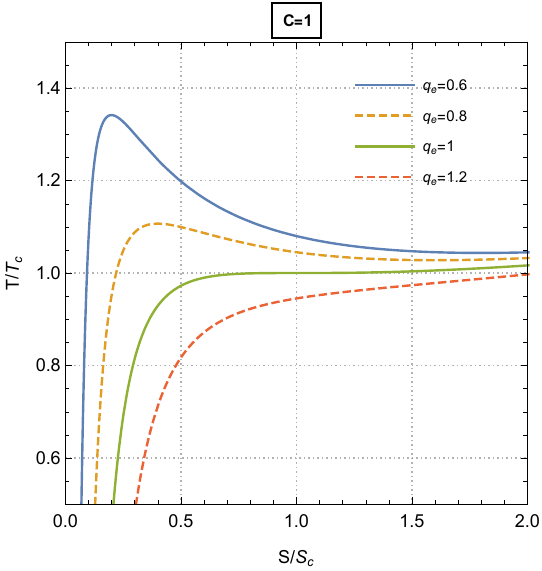}
    \end{subfigure}
    \hfill
    \begin{subfigure}[b]{0.4\textwidth}
        \includegraphics[scale=0.7]{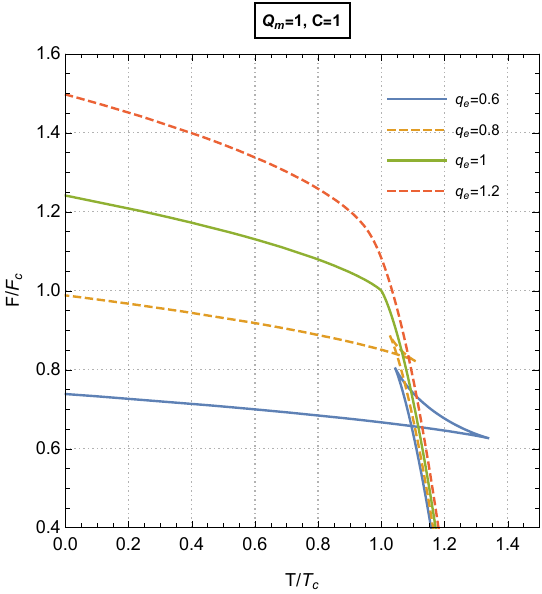}
    \end{subfigure}
    \caption{$T-S$ and $F-T$ plots for various values of $q_e$}
    \label{fig:two}
\end{figure}
We plot the $T-S$ plot and the $F-T$ plot using the equation given in \eqref{eq:TF}. The plot is seen in Figure \ref{fig:two}. We have already seen the plot in paper \cite{rps} so we will not go into detail explanation. Just would summarise that for the values of $\tilde{Q}_e$ on the range $0<\tilde{Q}_e<\tilde{Q}_e^c$, we see a change of phase of order one and it appears as a swallow tail in the $F-T$ plot. At $\tilde{Q}_e=\tilde{Q}_e^c$, that is at the critical value, we see that the order of the change of phase has been diverted to the order of two and it looks like a kink in $F-T$ plot. Above the critical points both the plots become monotonic. But the order has changed as for the dual charge the first order was seen for supercritical values of $\tilde{Q}_e$ whereas for a single charge it is seen at a subcritical value. We can also see the 3-D plot in Figure \ref{fig:3d2}.\\
Now when we set both the electrical and magnetic charges equal to zero, the black hole in the study reduces to Schwarzchild-AdS and we see the plot of $T-S$ and $F-T$ changes drastically. The temperature $T$ and the free energy $F$ at $\tilde{Q}_e=\tilde{Q}_m=0$ is:-
\begin{equation}
T=\frac{C \left(4 \pi  C S+3 S^2\right)}{8 \pi ^{3/2} l (C S)^{3/2}}, \quad F=\frac{S (4 \pi  C-S)}{8 \pi ^{3/2} l \sqrt{C S}}
\end{equation}
The critical points are:-
\begin{equation}
S_c=\frac{4C\pi}{3}, \quad T_c=\frac{\sqrt{3}}{2 l \pi}, \quad F_c=\frac{2 C}{3\sqrt{3} l}
\end{equation}
Then the EOS and the free energy can be written as:-
\begin{equation}
\label{eq:TF2}
t=\frac{ (s+1)}{2 \sqrt{ s}}, \quad f=\frac{(3-s) \sqrt{ s}}{2}
\end{equation}
where $t=T/T_c$, $f=F/F_c$ and $s=S/S_c$ are the relative parameters.
We plot the $T-S$ and $F-T$ plots using the equation \eqref{eq:TF2} and are seen in Figure \ref{fig:three}. It is noticed that the plots are quite different from the previous discussions and we can view only two branches where for the least entropy and small black hole we see an unstable branch and for the higher value of entropy and large black hole we notice a stable branch and in the $F-T$ plot we see the Hawking-Page transition \cite{rps}.
\\

\begin{figure}[htp]
    \centering
    \begin{subfigure}[b]{0.4\textwidth}
        \includegraphics[scale=0.7]{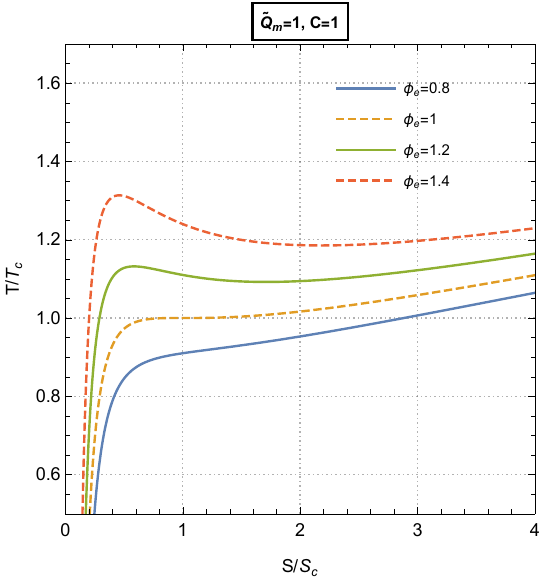}
    \end{subfigure}
    \hfill
    \begin{subfigure}[b]{0.4\textwidth}
        \includegraphics[scale=0.7]{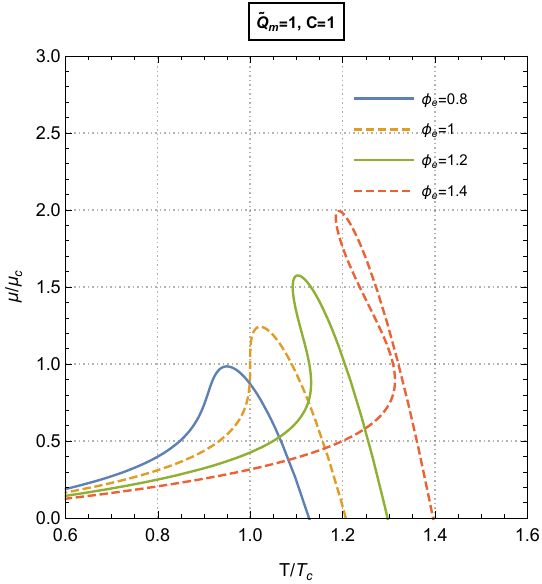}
    \end{subfigure}
    \caption{$T-S$ and $F-T$ plots for iso-voltage process}
    \label{fig:four}
\end{figure}

At a fixed $\tilde{\Phi}_e$ taking a mixed ensemble, we can also see the $T-S$ process, to obtain we use equation from \eqref{eq:cftall} and we get a temperature $T$ as a function $T(\tilde{\Phi}_e, \tilde{Q}_m,C,S)$ and is presented as:-
\begin{equation}
T=\frac{C \left(-4 \pi  C S \left(l^2 \tilde{\Phi}_e^2-1\right)-\pi ^2 \tilde{Q}_m^2+3 S^2\right)}{8 \pi ^{3/2} l (C S)^{3/2}}
\end{equation} Using equation \eqref{eq:differential}, we get the critical points as:-
\begin{equation}
S_c=\pi \tilde{Q}_m, \quad \tilde{\Phi}_e^c=\frac{\sqrt{2 C-3 \tilde{Q}_m}}{ \sqrt{2 C} l}, \quad T_c=\frac{\sqrt{\tilde{Q}_m}}{l \pi \sqrt{C}}
\end{equation}
Using the crucial values above we present the equation of state (EOS) as:-
\begin{equation}
t=\frac{ \sqrt{ \tilde{Q}_m} \left(\tilde{Q}_m \left(3 s^2+6 s \phi_e^2-1\right)-4 C s \left(\phi_e^2-1\right)\right)}{8 ( \tilde{Q}_m s)^{3/2}}
\end{equation}
where $t=T/T_c$, $s=S/S_c$ and $\phi_e=\tilde{\Phi}_e/\tilde{\Phi}_e^c$ are the required relative parameters.\\
For a clearer picture, we visualize the graph of $\mu-T$ where $\mu$ is the Gibbs free energy times $1/C$. Using equation \eqref{eq:cftall}, we can write $\mu$ in terms of $\tilde{\Phi}_e$ and introducing a relative parameter $\tilde{m}=\mu/\mu_c$, $s=S/S_c$ and $\phi_e=\tilde{\Phi}_e/\tilde{\Phi}_e^c$ so we get:-
\begin{equation}
\tilde{m}=-\frac{ \left(4 C s \left(\phi_e^2-1\right)+\tilde{Q}_m \left(s^2-6 s \phi_e^2+1\right)\right)}{4\tilde{Q}_m \sqrt{ s}}
\end{equation}
where $\mu_c=\frac{(C \tilde{Q}_m)^{3/2}}{2 C^3 l}$.
We plot the $T-S$ and $\mu-T$ for the various values of $\tilde{Q}_m$ seen in Figure \ref{fig:four}.\\

We see from figure \ref{fig:four} that for the subcritical value of $\tilde{\Phi}_e$, we see a monotonous curve in both the $T-S$ and $\mu-T$ plots and only two branches can be seen. But at the critical point, there is a change of phase and at supercritical values of the electrical potential, we see three phases for both the plots namely the lower, intermediate, and high entropy. In the $\mu-T$ plot for supercritical values, we see a turning of the plot (green, red). In the orange plot $(\phi_e=1)$, we see a change of phase of order two.\\
We also plot here the 3d plot for $T-S-\tilde{\Phi}_e$ just to get a view of the variation of all the variables combined.\\

\begin{figure}[htp]
\centering
\includegraphics[scale=0.7]{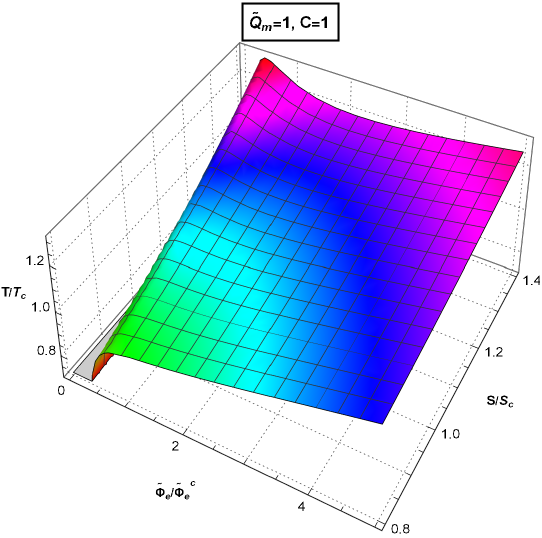}
\caption{$3d$ plot for $T-S-\tilde{\Phi}_e$}
\label{fig:five}
\end{figure}
\begin{figure}[htp]
    \centering
    \begin{subfigure}[b]{0.4\textwidth}
        \includegraphics[scale=0.7]{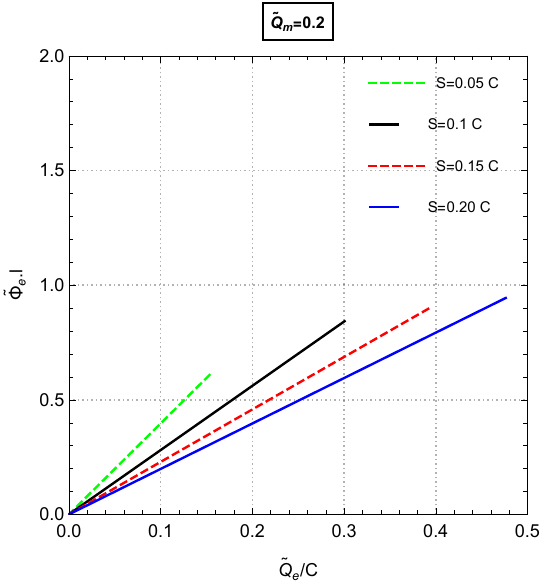}
    \end{subfigure}
    \hfill
    \begin{subfigure}[b]{0.4\textwidth}
        \includegraphics[scale=0.7]{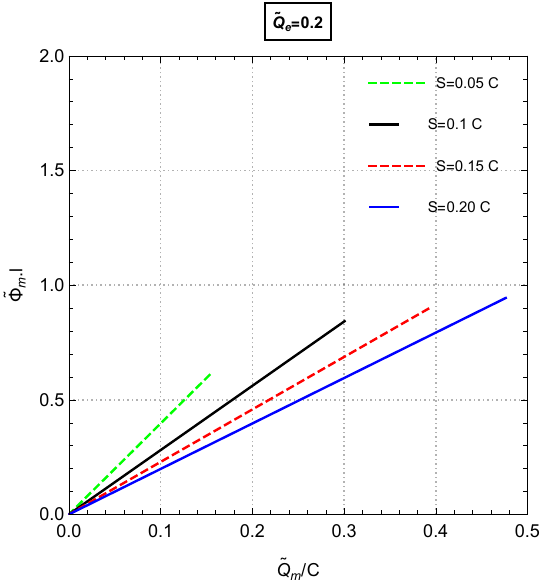}
    \end{subfigure}
    \caption{$\tilde{\Phi}_e-\tilde{Q}_e$ and $\tilde{\Phi}_m-\tilde{Q}_m$ plots}
    \label{fig:seven}
\end{figure}

At a fixed $\tilde{\Phi}_e$, $ \tilde{\Phi}_m$ we can also look into the $T-S$ process. We can obtain this expression for the equation of state by using the equation \eqref{eq:cftall}, and we obtain the temperature $T$ as
\begin{equation}
T=\frac{3 S-4 \pi  C \left(l^2 \left(\tilde{\Phi}_e^2+\tilde{\Phi}_m^2\right)-1\right)}{8 \pi ^{3/2} l \sqrt{C S}}
\end{equation}
From this equation, it is seen that there is a single point of extremity where the temperature becomes the minimum, the points are:-
\begin{equation}
S_{ex}=\frac{4}{3} \pi  C \left(1-l^2 \left(\tilde{\Phi}_e^2+\tilde{\Phi}_m^2\right)\right), \quad T_{ex}=\frac{\sqrt{3} \sqrt{ \left(1-l^2 \left(\tilde{\Phi}_e^2+\tilde{\Phi}_m^2\right)\right)}}{2 \pi l}
\end{equation}
By introducing some relative variables we can express the EOS explicitly depending on the entropy as:-
\begin{equation}
\label{eq:tphi}
t=\frac{s+1}{2 \sqrt{s}}
\end{equation}
where $s=S/S_{ex}$ and $t=T/T_{ex}$. As seen before, we plot the $\mu-T$ plot rather than the $F-T$ plot for a better understanding of the $T-S$ iso-voltage process. We see in the $\mu-T$ plot that the extremal value of the entropy $S=S_{ex}$ the chemical potential $\mu$ reaches the highest point and the value is:-
\begin{equation}
\mu_{ex}=\frac{2  \left(l^2 \left(\tilde{\Phi}_e^2+\tilde{\Phi}_m^2\right)-1\right)^2}{3 \sqrt{3} l \sqrt{ \left(1-l^2 \left(\tilde{\Phi}_e^2+\tilde{\Phi}_m ^2\right)\right)}}
\end{equation}
Introducing a relative variable $\tilde{m}=\frac{\mu}{\mu_{ex}}$, we can write the desired $\mu-T$ expression as:-
\begin{equation}
\label{eq:muphi}
\tilde{m}=\frac{(3-s) s}{2 \sqrt{s}}
\end{equation}
We can see that at no point in the equation \eqref{eq:tphi} and \eqref{eq:muphi} depends on the electric potential $\tilde{\Phi}_e$ and the magnetic potential $\tilde{\Phi}_m $ nor the central charge $C$. At an advanced stage, this can also be observed as the law of corresponding states. Using equations \eqref{eq:tphi} and \eqref{eq:muphi}, we can graph the $T-S$ and the $\mu-T$ plots and can be seen in Figure.
\begin{figure}[htp]
    \centering
    \begin{subfigure}[b]{0.4\textwidth}
        \includegraphics[scale=0.7]{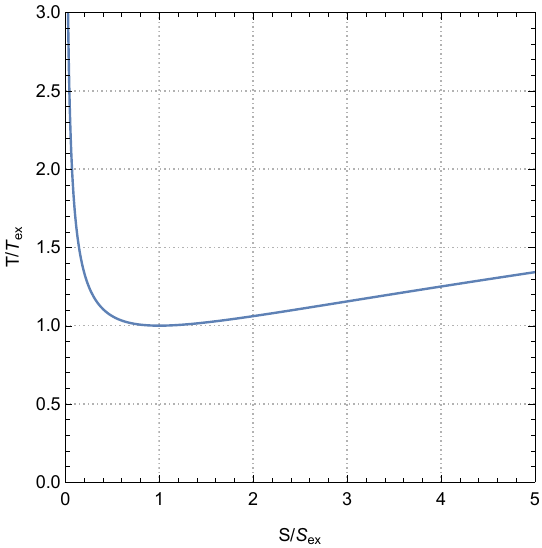}
    \end{subfigure}
    \hfill
    \begin{subfigure}[b]{0.4\textwidth}
        \includegraphics[scale=0.7]{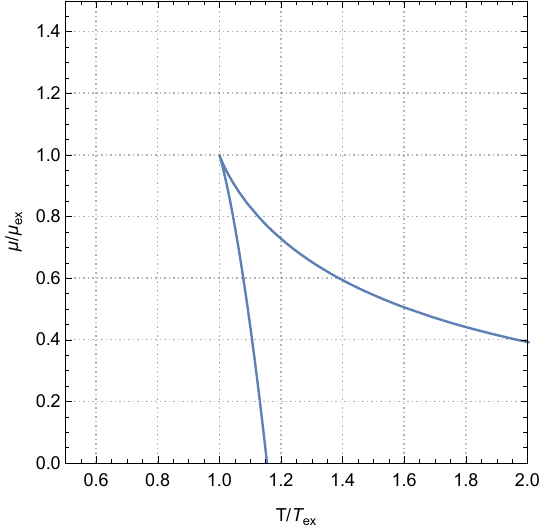}
    \end{subfigure}
    \caption{$T-S$ and $F-T$ plots}
    \label{fig:six}
\end{figure}
In the $T-S$ plot, we see transitions from the unstable small black hole to the stable black hole which can be related same with the plot where $\tilde{Q}_e=\tilde{Q}_m=0$ case (Schwarzschild black hole) in Figure \ref{fig:three}. In the $\mu-T$ plot, we see a Hawking-Page change of phase which is seen at some temperature $T_{HP}$ shown as $\mu (T_{HP},\tilde{\Phi}_e, \tilde{\Phi}_m)=0$. Using equation \eqref{eq:muphi} we get $s=3$ and by putting it back in the equation \eqref{eq:tphi} we get $T_{HP}=\frac{2}{\sqrt{3}}T_{ex}$.\\
We can also see the thermodynamic process for the $\tilde{\Phi}_e-\tilde{Q}_e$ and $\tilde{\Phi}_m-\tilde{Q}_m$ curves. But if we look closely at the equation \eqref{eq:cftall}, we can notice that both the electric and magnetic potential $\tilde{\Phi}_e$, $\tilde{\Phi}_m$ are directly proportional to the electric charge $\tilde{Q}_e$ and magnetic charge $\tilde{Q}_m$ respectively. So, the adiabatic iso-C-charge process for the $\tilde{\Phi}_e-\tilde{Q}_e$ and $\tilde{\Phi}_m-\tilde{Q}_m$ plots does not show any points of inflection nor points of extremity. We can  see the plots in Figure \ref{fig:seven}.\\
For the $\mu-C$ process, rather than all the other processes examined above, we see that the $\mu-C$ process calls for the dependency on the change of the central charge $C$.\\
We start from the equation \eqref{eq:cftall} and obtain the equation of $\mu$ for which the $\mu-C$ process can be obtained. When we graph the $\mu-C$ plot, it is seen that there is a point of extremity in the plot for the fixed values of $(S, \tilde{Q}_e, \tilde{Q}_m)$. Therefore the point of extremity is obtained as:-
\begin{equation}
C_{max}= \frac{3 \left(\pi ^2 \tilde{Q}_e^2+\pi ^2 \tilde{Q}_m^2+S^2\right)}{4 \pi  S}, \quad \mu_{max}=\frac{2 S}{3 \sqrt{3} l \sqrt{\pi ^2 \left(\tilde{Q}_e^2+\tilde{Q}_m^2\right)+S^2}}
\end{equation}
\begin{figure}
\centering
\includegraphics[scale=0.8]{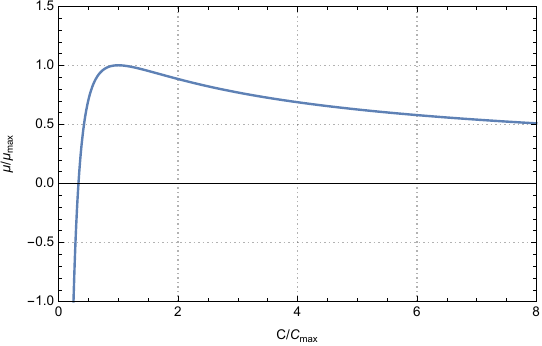}
\caption{$\mu-C$ process}
\label{fig:eight}
\end{figure}
We give in certain dimensionless parameters as in:-
\begin{equation}
c=\frac{C}{C_{max}}, \quad m=\frac{\mu}{\mu_{max}}
\end{equation}
By using this parameters, the $\mu$ equation becomes as:-
\begin{equation}
\label{eq:muc}
m=\frac{3 c-1}{2 c^{3/2}}
\end{equation}
We can see from the plot that for any fixed values of the variable $(\tilde{Q}_e, \tilde{Q}_m, S)$, we get the same $\mu-C$ plot and also, this plot appears to the same when this process was calculated for the RN-AdS and Kerr AdS  black holes which emphasize about a universality behind this $\mu-C$ process.
From the equation \eqref{eq:muc}, we see that $\mu$ becomes zero at $C=\frac{1}{3}C_{max}$ where it can already be seen in Figure \ref{fig:eight} about where it cuts the $C$ axis.
\section{Equation of states and homogeneity using the R\'enyi entropy}
\label{sec:four}
The R\'enyi entropy is given as:- 
\begin{equation}
S=\frac{1}{\lambda}\log[1+\lambda\left(\pi \frac{r^2}{G}\right)]
\end{equation}
where $\lambda$ is the R\'enyi parameter or order of entropy.
Solving this entropy, we get the value of $r$ as:-
\begin{equation}
r=\frac{\sqrt{G} \sqrt{e^{\lambda  S}-1}}{\sqrt{\pi } \sqrt{\lambda }}
\end{equation}
Putting this value if $r$ in the mass equation \eqref{eq:mass1} we get:-
\begin{equation}
M=\frac{G \left(e^{\lambda  S}-1\right)^2+\pi  \lambda  l^2 \left(\pi  \lambda  \left(\tilde{Q}_e^2+\tilde{Q}_m^2\right)+e^{\lambda  S}-1\right)}{2 \pi ^{3/2} \sqrt{G} \lambda ^{3/2} l^2 \sqrt{e^{\lambda  S}-1}}
\end{equation}
We substitute the electric and magnetic charge with the rescaled dual charges from equation \eqref{eq:rcharge1}, \eqref{eq:rcharge2} we get:-
\begin{equation}
M=\frac{G \left(\pi ^2 \lambda ^2 \left(\tilde{Q}_e^2+\tilde{Q}_m^2\right)-2 e^{\lambda  S}+e^{2 \lambda  S}+1\right)+\pi  \lambda  l^2 \left(e^{\lambda  S}-1\right)}{2 \pi ^{3/2} \sqrt{G} \lambda ^{3/2} l^2 \sqrt{e^{\lambda  S}-1}}
\end{equation}
Putting the value of the Newton's constant $G_N$  from equation \eqref{eq:mu}, we get:-
\begin{equation}
M=\frac{\sqrt{\frac{l^2}{C}} \left(\pi  \lambda  \left(\pi  \lambda  \left(\tilde{Q}_e^2+\tilde{Q}_m^2\right)-C\right)+(\pi  C \lambda -2) e^{\lambda  S}+e^{2 \lambda  S}+1\right)}{2 \pi ^{3/2} \lambda ^{3/2} l^2 \sqrt{e^{\lambda  S}-1}}
\end{equation}
As the study gets very complicated we use the about equation so in simplification we series expand for a small value of $\lambda$. Hence, we get the value of mass M as:-
\begin{equation}
\label{eq:cftmass1}
M=\frac{\sqrt{\frac{l^2}{C}} \left(\pi  C S (\lambda  S+4)-\pi ^2 \left(\tilde{Q}_e^2+\tilde{Q}_m^2\right) (\lambda  S-4)+S^2 (3 \lambda  S+4)\right)}{8 \pi ^{3/2} l^2 \sqrt{S}}
\end{equation}
Using the first law given in equation \eqref{eq:first}, we derive the various thermodynamic variables as:-
\begin{equation}
\label{eq:cftall1}
\begin{split}
& T=\frac{\sqrt{\frac{l^2}{C}} \left(\pi  C S (3 \lambda +4)-\pi ^2 \left(\tilde{Q}_e^2+\tilde{Q}_m^2\right) (\lambda  S+4)+3 S^2 (5 \lambda  S+4)\right)}{16 \pi ^{3/2} l^2 S^{3/2}}\\
& \tilde{\Phi}_e= -\frac{\sqrt{\pi } \tilde{Q}_e \sqrt{\frac{l^2}{C}} (\lambda S-4)}{4 l^2 \sqrt{S}}\\
& \tilde{\Phi}_m= -\frac{\sqrt{\pi } \tilde{Q}_m \sqrt{\frac{l^2}{C}} (\lambda  S-4)}{4 l^2 \sqrt{S}}\\
& \mu=\frac{\pi  C S (\lambda  S+4)+\pi ^2 \left(\tilde{Q}_e^2+\tilde{Q}_m^2\right) (\lambda S-4)-\left(S^2 (3 \lambda S+4)\right)}{16 \pi ^{3/2} C^2 \sqrt{S} \sqrt{\frac{l^2}{C}}}
\end{split}
\end{equation}
We can also check the homogeneity for the mass equation and the rest simply by rescaling $S \rightarrow\Lambda S$, $\tilde{Q}_e \rightarrow \Lambda \tilde{Q}_e$, $ \tilde{Q}_m\rightarrow\Lambda\tilde{Q}_m$, $C\rightarrow\Lambda C$ and $\lambda\rightarrow\frac{\lambda}{\Lambda}$. Putting this in equation \eqref{eq:cftmass1}, \eqref{eq:cftall1}, we get:-
\begin{equation}
\begin{split}
& M=\frac{\Lambda \left(\pi  C S (\lambda  S+4)-\pi ^2 \left(\tilde{Q}_e^2+\tilde{Q}_m^2\right) (\lambda  S-4)+S^2 (3 \lambda  S+4)\right)}{8 \pi ^{3/2} C \sqrt{\Lambda S} \sqrt{\frac{l^2}{\Lambda C}}}=\Lambda^1 M\\
& T=\frac{\Lambda \left(\pi  C S (3 \lambda  S+4)-\pi ^2 \left(\tilde{Q}_e^2+\tilde{Q}_m^2\right) (\lambda  S+4)+3 S^2 (5 \lambda  S+4)\right)}{16 \pi ^{3/2} C (\Lambda S)^{3/2} \sqrt{\frac{l^2}{\Lambda C}}}=\Lambda^0 T
\end{split}
\end{equation}
We note that the mass is of order one where the temperature and the rest variables when checked are of order zero.
\subsection{Thermodynamic processes and phase transitions}
\begin{figure}[htp]
    \centering
    \begin{subfigure}[b]{0.4\textwidth}
        \includegraphics[scale=0.7]{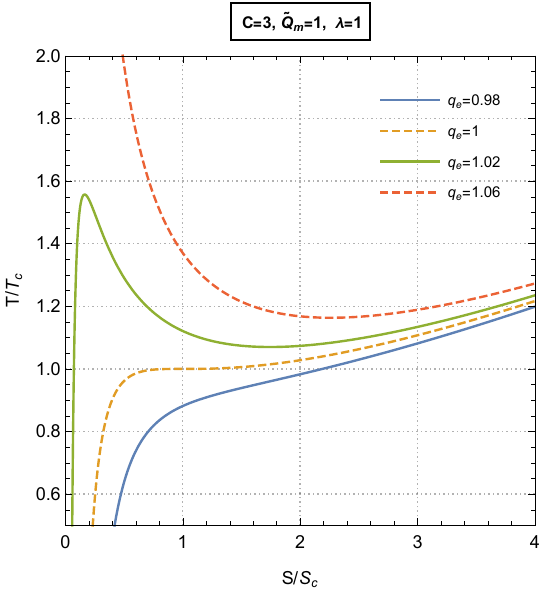}
        \caption{}
        \label{fig:sub1}
    \end{subfigure}
    \hfill
    \begin{subfigure}[b]{0.4\textwidth}
        \includegraphics[scale=0.7]{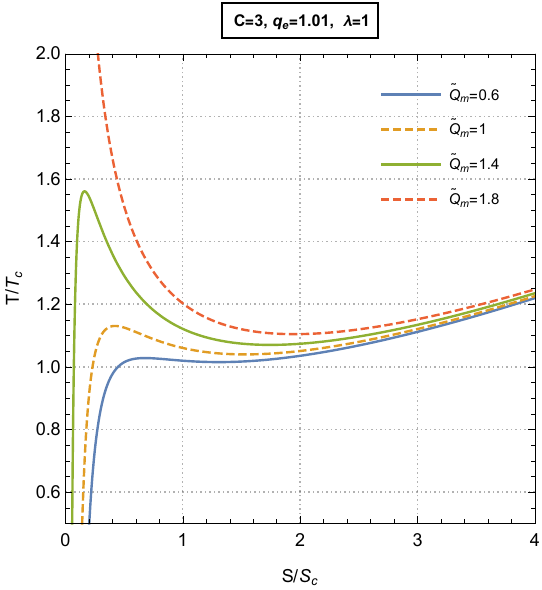}
        \caption{}
        \label{fig:sub2}
    \end{subfigure}
    \begin{subfigure}[b]{0.4\textwidth}
        \includegraphics[scale=0.7]{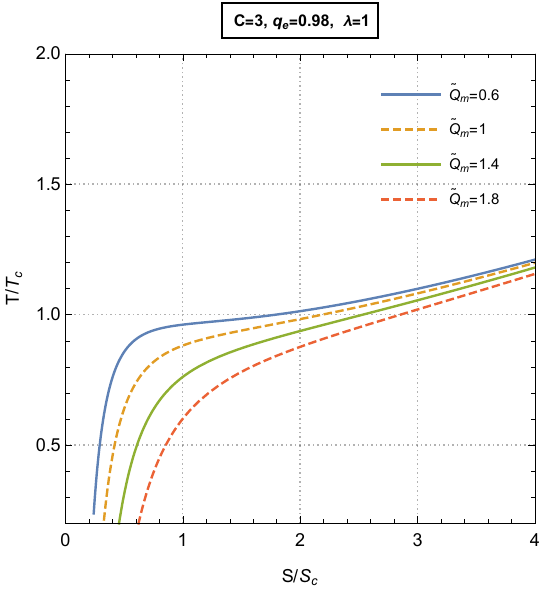}
        \caption{}
        \label{fig:sub3}
    \end{subfigure}
    \hfill
    \begin{subfigure}[b]{0.4\textwidth}
        \includegraphics[scale=0.7]{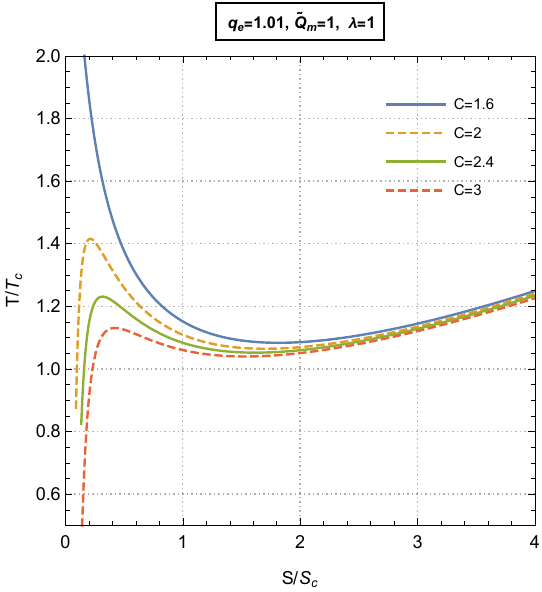}
        \caption{}
        \label{fig:sub4}
    \end{subfigure}
    \begin{subfigure}[b]{0.4\textwidth}
        \includegraphics[scale=0.7]{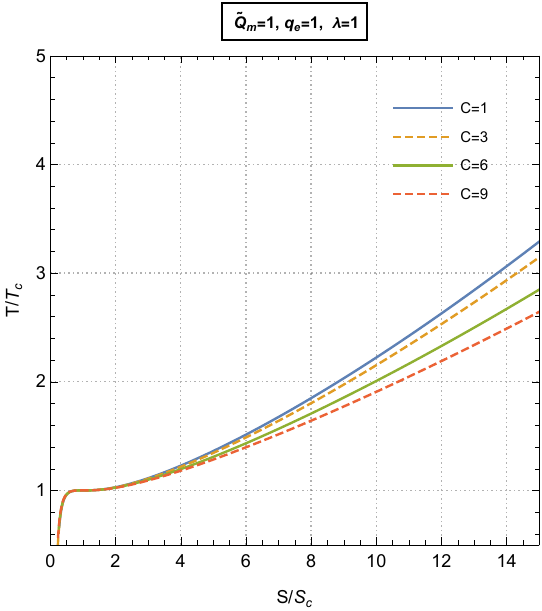}
        \caption{}
        \label{fig:sub5}
    \end{subfigure}
    \hfill
    \begin{subfigure}[b]{0.4\textwidth}
        \includegraphics[scale=0.7]{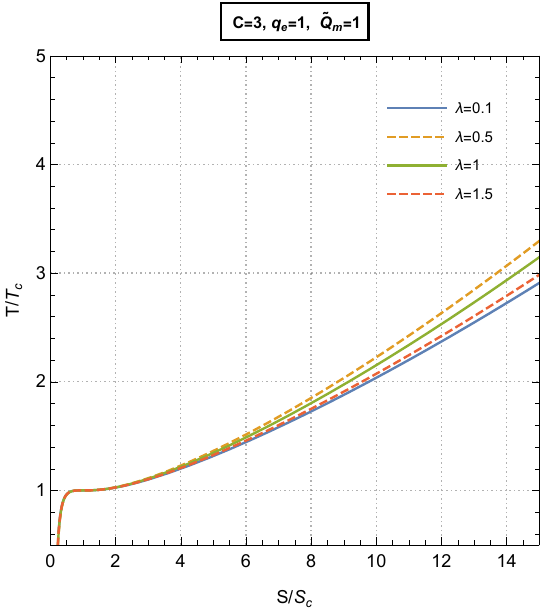}
        \caption{}
        \label{fig:sub6}
    \end{subfigure}
    \caption{$T-S$ plots}
    \label{fig:nine}
\end{figure}

We study the different thermodynamic processes for the dual-charged AdS black hole in the Restricted Phase Space Thermodynamics using the R\'enyi entropy.\\
When we put $T=0$, then the upper bound for the electric charge $\tilde{Q}_e$ for various values of $\tilde{Q}_m$, $S$ and $C$ is written as:-
\begin{equation}
\tilde{Q}_e^{max}=\frac{\sqrt{-\pi  C S (3 \lambda  S+4)+\pi ^2 \tilde{Q}_m^2 (\lambda  +4)-3 S^2 (5 \lambda +4)}}{\pi  \sqrt{-\lambda  S-4}}
\end{equation}
We find the point of inflection by using equation \eqref{eq:differential} for the T-S plots and hence the electric charge $\tilde{Q}_e$ takes the critical value as:-
\begin{equation}
\label{eq:critq}
\tilde{Q}_e^c=\frac{\sqrt{-\pi ^2 \left(\frac{A_1}{90 \lambda }+12\right) \tilde{Q}_m^2-\frac{\left(A_1\right)^2 \left(\frac{A_1}{6 \lambda }+4\right)}{2700 \lambda ^4}+\frac{C \pi  \left(A_1\right) \left(4-\frac{A_1}{30 \lambda }\right)}{90 \lambda ^2}}}{\pi  \sqrt{\frac{A}{90 \lambda }+12}}
\end{equation}
where the values of $A_1$ are given in Appendix \ref{appendix}.
Where by putting various values of $C$, $\tilde{Q}_m$ to get the critical electric charge. The critical value of $S$ and $T$ can be given as:-
\begin{equation}
\label{eq:critc}
S_c=\frac{A_2+\frac{\lambda ^2 (\pi  C \lambda -536)^2}{A_2}-\lambda  (\pi  C \lambda +544)}{90 \lambda ^2}
\end{equation}
where the value of $A_2$ given in Appendix \ref{appendix}
and
\begin{equation}
\label{eq:critt}
T_c=\frac{135 \sqrt{\frac{5}{2}} \sqrt{\frac{l^2}{C}} \left(\frac{\left(A_1\right)^2 \left(\frac{A_1}{18 \lambda }+4\right)}{2700 \lambda ^4}+\frac{C \pi  \left(A_1\right) \left(\frac{A_1}{30 \lambda }+4\right)}{90 \lambda ^2}-\pi ^2 \left(\frac{A_1}{90 \lambda }+4\right) \left(\tilde{Q}_m^2+A_4\right)\right)}{4 l^2 \pi ^{3/2} \left(\frac{A_1}{\lambda ^2}\right)^{3/2}}
\end{equation}
where the values of $A_1$, $A_4$ are in Appendix \ref{appendix}.
We introduce some relational variables
\begin{equation}
t=\frac{T}{T_c}, \quad s=\frac{S}{S_c}, \quad q_e=\frac{\tilde{Q_e}}{\tilde{Q}_e^c}
\end{equation}
Using this variables and the critical values from \eqref{eq:critq}, \eqref{eq:critc}, \eqref{eq:critt} and putting in equation \eqref{eq:cftall1} we get the equation of state as:-
\begin{equation}
\label{eq:tlam}
t=\frac{ \left(\frac{s^2 \left(A_1\right)^2 \left(\frac{s \left(A_1\right)}{18 \lambda }+4\right)}{2700 \lambda ^4}+\frac{C \pi  s \left(A_1\right) \left(\frac{s \left(A_1\right)}{30 \lambda }+4\right)}{90 \lambda ^2}-\pi ^2 \left(\frac{s \left(A_1\right)}{90 \lambda }+4\right) \left(\tilde{Q}_m^2+q_e^2 A_4\right)\right)}{\left(s\right)^{3/2} \left(\frac{\left(A_1\right)^2 \left(\frac{A_1}{18 \lambda }+4\right)}{2700 \lambda ^4}+\frac{C \pi  \left(A_1\right) \left(\frac{A_1}{30 \lambda }+4\right)}{90 \lambda ^2}-\pi ^2 \left(\frac{A_1}{90 \lambda }+4\right) \left(\tilde{Q}_m^2+A_4\right)\right)}
\end{equation}
where the values of $A_1$, $A_4$ are given in Appendix \ref{appendix}.
Using equation \eqref{eq:cftmass1} and \eqref{eq:cftall1}, the helmholtz free energy is given by:-
\begin{equation}
F(T,\tilde{Q}_e, \tilde{Q}_m, C)= M(S,\tilde{Q}_e, \tilde{Q}_m,C)-TS
\end{equation}
By using the critical values from \eqref{eq:critq}, \eqref{eq:critc}, \eqref{eq:critt} we get the critical free energy as:-
\begin{equation}
\label{eq:critf}
\resizebox{1\hsize}{!}{$
F_c=-\frac{ \sqrt{\frac{A_1}{\lambda ^2}} \left(C \pi  \lambda  \left(-360 \lambda + A_1\right)+\left(A_1\right) \left(\frac{A_1}{10 \lambda }+4\right)+\frac{3 \left(\frac{A_1}{90 \lambda }-12\right) \left(C \pi  \lambda  \left(+120 \lambda -A_1\right)-\left(A_1\right) \left(\frac{A_1}{6 \lambda }+4\right)\right)}{\frac{A_1}{90 \lambda }+12}\right)}{4320 \sqrt{10} l \pi ^{3/2} C^{1/2} \lambda ^2}$}
\end{equation}
value of $A_1$ given in Appendix \ref{appendix}.
Introducing a relative variable $f=\frac{F}{F_c}$, we can write the free energy equation using all the equations \eqref{eq:critq}, \eqref{eq:critc}, \eqref{eq:critt}, \eqref{eq:critf} as:-
\begin{equation}
\label{eq:flam}
f=\frac{8100 \lambda ^2 \left(\frac{s^2 A_1^2 \left(\frac{s \left(A_1\right)}{10 \lambda }+4\right)}{8100 \lambda ^4}+\frac{C \pi  s \left(A_1\right) \left(\frac{s \left(A_1\right)}{90 \lambda }-4\right)}{90 \lambda ^2}+\pi ^2 \left(\frac{s \left(A_1\right)}{90 \lambda }-12\right) \left(\tilde{Q}_m^2+A_4\right)\right)}{\frac{A_1}{\lambda ^2} \sqrt{s} \left(C \pi  \lambda  \left(-360 \lambda + A_1\right)+\left(A_1\right) \left(\frac{A_1}{10 \lambda }+4\right)+A_5\right)}
\end{equation}
vaules of $A_1$, $A_4$, $A_5$ in given in Appendix \ref{appendix}.
We can now plot the $T-S$ seen in Figure \ref{fig:nine}  relating to the iso-e-charge process and other various fixed values to make a clear picture of various factor dependencies using equation \eqref{eq:tlam}.

\begin{figure}[htp]
    \centering
    \begin{subfigure}[b]{0.4\textwidth}
        \includegraphics[scale=0.7]{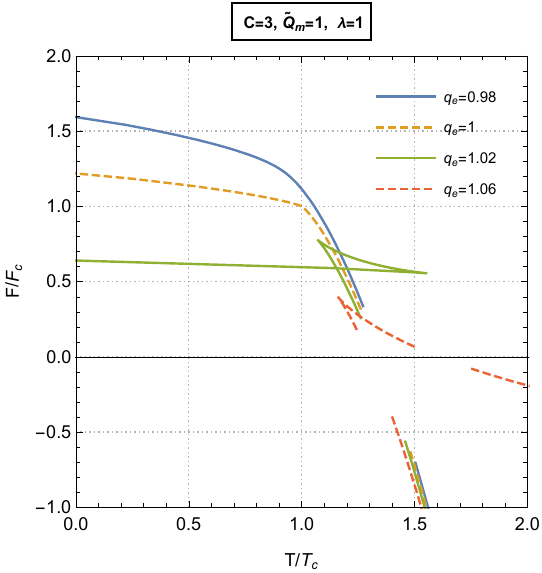}
        \caption{}
        \label{fig:sub11}
    \end{subfigure}
    \hfill
    \begin{subfigure}[b]{0.4\textwidth}
        \includegraphics[scale=0.7]{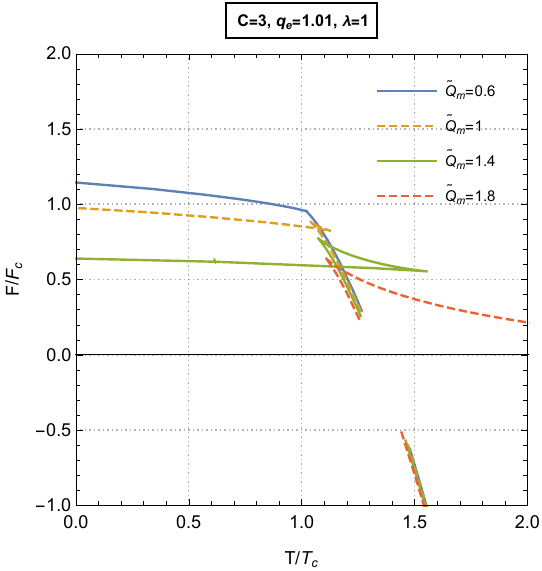}
        \caption{}
        \label{fig:sub12}
    \end{subfigure}
    \begin{subfigure}[b]{0.4\textwidth}
        \includegraphics[scale=0.7]{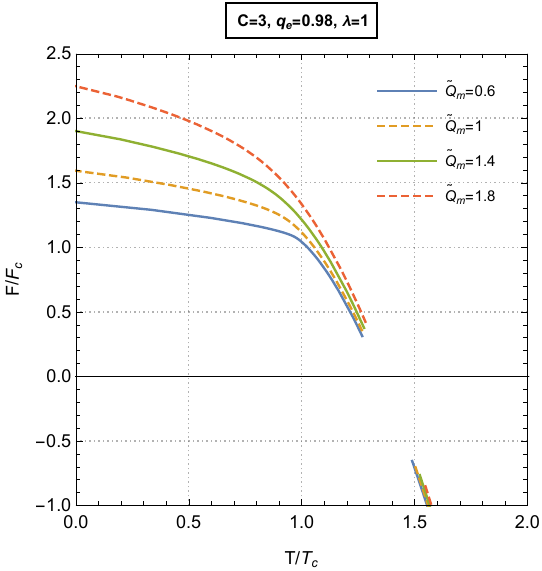}
        \caption{}
        \label{fig:sub13}
    \end{subfigure}
    \hfill
    \begin{subfigure}[b]{0.4\textwidth}
        \includegraphics[scale=0.7]{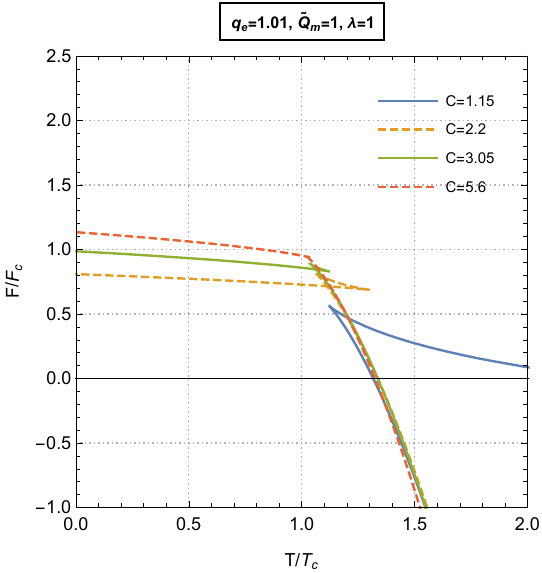}
        \caption{}
        \label{fig:sub14}
    \end{subfigure}
    \caption{$F-T$ plots}
    \label{fig:ten}
\end{figure}

\begin{figure}[htp]
    \centering
    \begin{subfigure}[b]{0.4\textwidth}
        \includegraphics[scale=0.7]{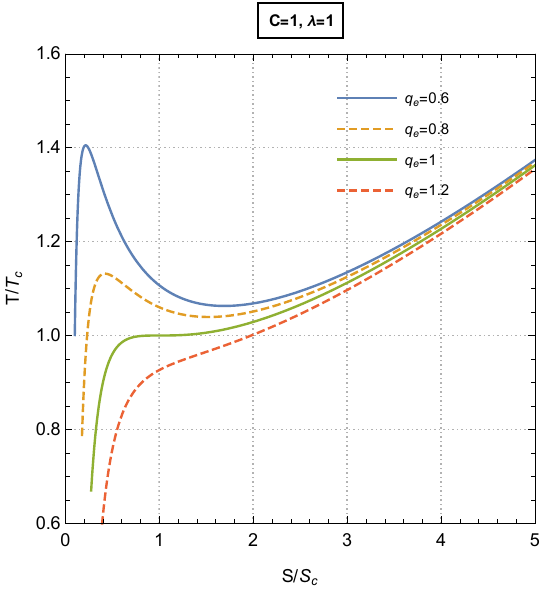}
        \caption{}
        \label{fig:sub21}
    \end{subfigure}
    \hfill
    \begin{subfigure}[b]{0.4\textwidth}
        \includegraphics[scale=0.7]{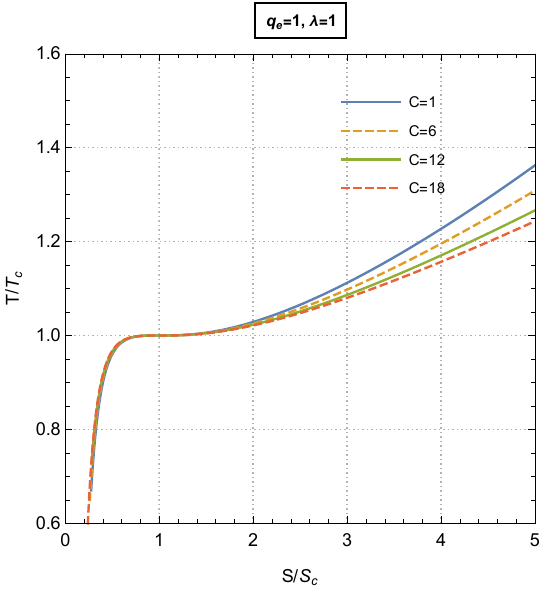}
        \caption{}
        \label{fig:sub22}
    \end{subfigure}
    \begin{subfigure}[b]{0.4\textwidth}
        \includegraphics[scale=0.7]{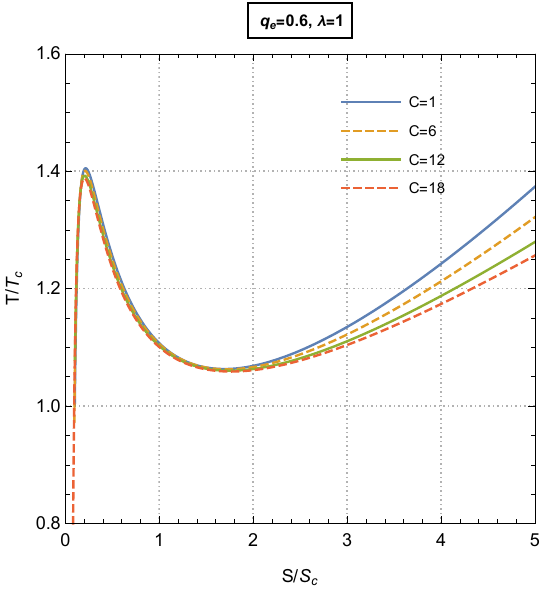}
        \caption{}
        \label{fig:sub23}
    \end{subfigure}
    \hfill
    \begin{subfigure}[b]{0.4\textwidth}
        \includegraphics[scale=0.7]{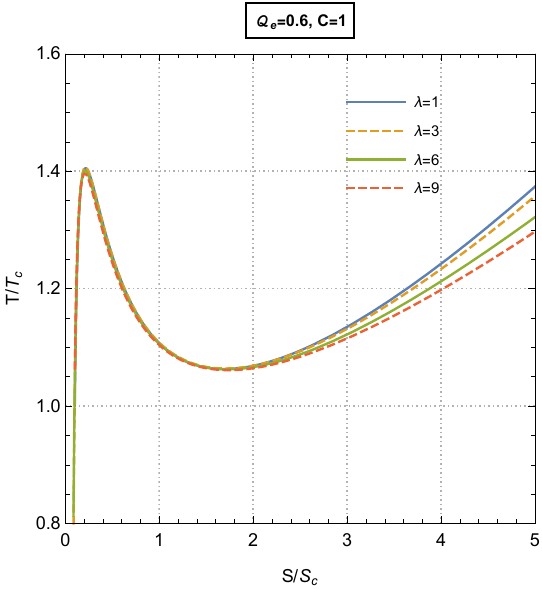}
        \caption{}
        \label{fig:sub24}
    \end{subfigure}
    \caption{$T-S$ plots}
    \label{fig:eleven}
\end{figure}

We see in the plot (\ref{fig:sub1}) below the crucial point of $q_e$, the plot is monotonically increasing, which is also similar to the $F-T$ plot. Now when it has a value more than the critical value of $q_e$, the curve becomes non-monotonic and we see a swallowtail-like structure in the $F-T$ graph. This kind of structure is the Van der Waals type phase transition of the first order of the iso-$e$-charge processes. At $\tilde{Q}_e=\tilde{Q}_e^c$, we observe a change of phase of order two. If we observe the red plot in Fig \ref{fig:sub1}, we see as the temperature decreases in value, the entropy increases till it reaches a stable value and goes under a non-equilibrium phase transition from an unstable small black hole to a stable large black hole. Similarly, in the $F-T$ plot in Figure \ref{fig:sub11}, for the same value of $\tilde{Q}_e$, we see the Hawking-type phase transition as seen for the red line.\\
In figure \ref{fig:sub2} \ref{fig:sub12}, we have plotted the $T-S$ and $F-T$ plot by using a supercritical value of $\tilde{Q}_e$ we plot for various values of $\tilde{Q}_m$. We notice that in both the plots we see a first-order phase transition, especially in $F-T$ it is like a swallow tail and for certain high values of $\tilde{Q}_m$, we see Hawking-Page type phase transition (red plot).\\
In Figure \ref{fig:sub3}, \ref{fig:sub13} we set for a subcritical value of the electric charge $\tilde{Q}_e$ and plot for various values of the magnetic charge $\tilde{Q}_m$. We note that all the different values of $\tilde{Q}_m$ plots are monotonous and have no change in phase. This same can be seen in the $F-T$ plot.\\
Now for the Figure \ref{fig:sub4}, \ref{fig:sub14} we fix a supercritical value of $\tilde{Q}_e$ and other variables and plot for various values of the CFT's central charge $C$. Here in the $T-S$ and $F-T$ plot one thing to be noticed is that as the value of C increases, it shifts from the Hawking's-Page type phase transition to the Van der Waals type phase transition of order one. This is opposite when we change the value of $\tilde{Q}_e$ and $\tilde{Q}_m$, as the value increases, like in Figure (\ref{fig:sub1}), it shifts from Vander-Waals type phase transition then to Hawking's-Page type.\\
In Figure \ref{fig:sub5}, for the critical value $\tilde{Q}_e$, we plot for various values of central charge and we see that at a critical value of $\tilde{Q_e}$, we get change of phase of order two and for various values of entropy, we do not see any change for low values of entropy but for large values of entropy, we see that the plots diverge and for a particular $S$ value, the higher the value of $C$, the less the value of $T$. For Figure \ref{fig:sub6}, we plot for a critical value of $\tilde{Q}_e$ and other variables setting constant and varying $\Lambda$. We see that as the value of $\Lambda$ increases, we do not see any difference in the low entropy plots but do see some distinction in the high entropy range also one thing to be noticed is that the variation of $\lambda$ matches with the change seen with the change in central charge plot $C$. 
\\ Now we can also do some study by setting the magnetic charge $\tilde{Q}_m=0$. By setting this condition, the $T-S$ and $F-T$ plots become a bit different by being devoid of the non-equilibrium in the $T-S$ and Hawking-Page in the $F-T$ plot. Setting $\tilde{Q}_m=0$ in \eqref{eq:cftall1}, we calculate the temperature as:-
\begin{equation}
T=\frac{ \left(\pi  C S (3 \lambda  S+4)-\pi ^2 \tilde{Q}_e^2 (\lambda  S+4)+3 S^2 (5 \lambda  S+4)\right)}{16 \pi ^{3/2} C^{1/2} l S^{3/2}}
\end{equation}
Now using the equation \eqref{eq:differential}, we can calculate the critical values of the entropy $S$ and the electric charge $\tilde{Q}_e$ as:-
\begin{equation}
\label{eq:scc}
\begin{split}
& S_c=\frac{\sqrt[3]{B_1}+\frac{\lambda ^2 (\pi  C \lambda -536)^2}{\sqrt[3]{B_1}}-\lambda  (\pi  C \lambda +544)}{90 \lambda ^2}\\
\end{split}
\end{equation}
where $B_1$ is given in Appendix \ref{appendix} and
\begin{equation}
\label{eq:qccc}
\begin{split}
& \tilde{Q}_e^c=\frac{\sqrt{\frac{C \pi  \left(4-\frac{B_3}{30 \lambda }\right)-\frac{\left(B_3\right) \left(\frac{B_3}{6 \lambda }+4\right)}{30 \lambda ^2}}{\left(\frac{B_3}{\lambda ^2}\right)^{3/2}}}}{3 \sqrt{10} \pi  \sqrt{\frac{\frac{B_3}{90 \lambda }+12}{\left(\frac{B_3}{\lambda ^2}\right)^{5/2}}}}\\
\end{split}
\end{equation}
Using the critical values from \eqref{eq:scc}, \eqref{eq:qccc} we can get the critical values for the temperature $T$ and the free energy $F$ as:-
\begin{equation}
\label{eq:tcc}
T_c=\frac{ \left(C \pi  \left(120 \lambda+ B_3\right) \lambda +B_3 \left(\frac{B_3}{18 \lambda }+4\right)-\frac{B_4}{\frac{B_3}{90 \lambda }+12}\right)}{16 \sqrt{10 C} l \pi ^{3/2} \lambda ^2 \sqrt{\frac{B_3}{\lambda ^2}}}
\end{equation}
\begin{equation}
 F_c=-\frac{ \sqrt{\frac{B_3}{\lambda ^2}} \left(C \pi  \left(-360 \lambda +B_3\right) \lambda +B_3 \left(\frac{B_3}{10 \lambda }+4\right)+\frac{B_4}{\frac{B_3}{90 \lambda }+12}\right)}{4320 \sqrt{10 C} l \pi ^{3/2} \lambda ^2}
\end{equation}
where $B_1$, $B_2$, $B_3$, $B_4$ are given in Appendix \ref{appendix}
where the free energy $F$ is given as:-
\begin{equation}
F=-\frac{\sqrt{\frac{l^2}{C}} \left(\pi  C S (\lambda  S-4)+\pi ^2 \tilde{Q}_e^2 (\lambda  S-12)+S^2 (9 \lambda  S+4)\right)}{16 \pi ^{3/2} l^2 \sqrt{S}}
\end{equation}
Using the critical equations \eqref{eq:scc}, \eqref{eq:qccc}, \eqref{eq:tcc} we can write the equation of state and the free energy as:-
\begin{equation}
t=\frac{\left(\frac{B_3}{\lambda ^2}\right)^{3/2} \left(B_3 \left(\frac{s B_3}{18 \lambda }+4\right) s^2+C \pi  \lambda  \left(120 \lambda +s B_3\right) s-\frac{30 q_e^2 \lambda ^2 \left(\frac{s B_3}{90 \lambda }+4\right) B_4}{\frac{B_3}{90 \lambda }+12}\right)}{\left(\frac{s B_3}{\lambda ^2}\right)^{3/2} \left(C \pi  B_3 \lambda +B_3 \left(\frac{B_3}{18 \lambda }+4\right)-\frac{30 \lambda ^2 \left(\frac{B_3}{90 \lambda }+4\right) B_4}{\frac{B_3}{90 \lambda }+12}\right)}
\end{equation}
\begin{equation}
f=\frac{ \left(B_3 \left(\frac{s B_3}{10 \lambda }+4\right) s^2+C \pi  \lambda  \left(s B_3-360 \lambda \right) s+\frac{90 q_e^2 \lambda ^2 \left(\frac{s B_3}{90 \lambda }-12\right) B_4}{\frac{B_3}{90 \lambda }+12}\right)}{\sqrt{s} \left(C \pi  \left(B_3-360 \lambda \right) \lambda +B_3 \left(\frac{B_3}{10 \lambda }+4\right)+\frac{90 \lambda ^2 \left(\frac{B_3}{90 \lambda }-12\right) B_4}{\frac{B_3}{90 \lambda }+12}\right)}
\end{equation}
where $t=\frac{T}{T_c}$, $q_e=\frac{\tilde{Q}_e}{\tilde{Q}_e^c}$, $s=\frac{S}{S_c}$ and $f=\frac{F}{F_c}$. We plot the $T-S$ and $F-T$ plot using equations and shown in Figure \ref{fig:eleven}.

\begin{figure}[htp]
    \centering
    \begin{subfigure}[b]{0.4\textwidth}
        \includegraphics[scale=0.7]{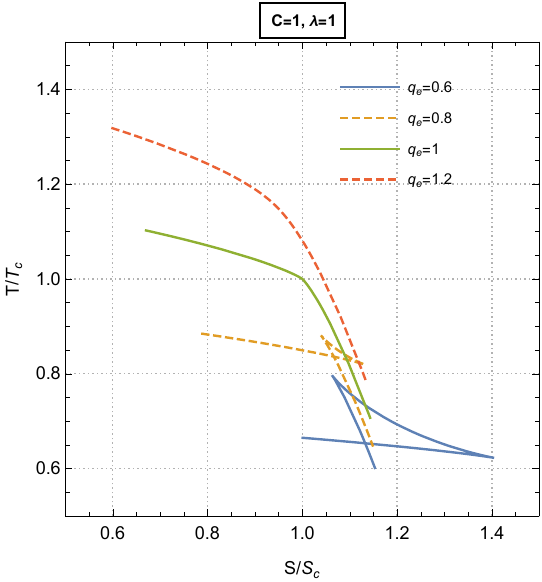}
        \caption{}
        \label{fig:sub25}
    \end{subfigure}
    \hfill
    \begin{subfigure}[b]{0.4\textwidth}
        \includegraphics[scale=0.7]{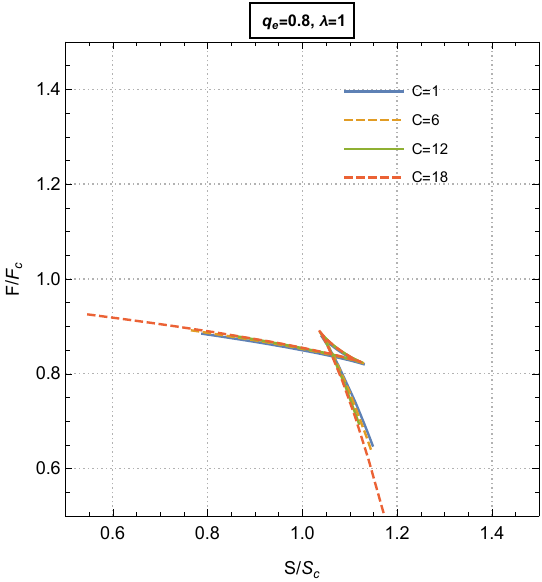}
        \caption{}
        \label{fig:sub26}
    \end{subfigure}
    \caption{$F-T$ plots}
    \label{fig:twelve}
\end{figure}
In Figure \ref{fig:sub21}, \ref{fig:sub25}, we see the $T-S$ and $F-T$ plot primarily show Van der Waals type phase transition and in the range $0<\tilde{Q}_e<\tilde{Q}_e^c$, we see a first-order phase transition and in the $F-T$ plot we see a swallow-tail like structure and at $\tilde{Q}_e=\tilde{Q}_e^c$ we get the second order phase transition and in the $F-T$ plot we get a kink like structure. For the supercritical values of $\tilde{Q}_e$, we have monotonically increasing plots in both Figure \ref{fig:sub21}, \ref{fig:sub25}. \\
Now for Figure \ref{fig:sub23}, \ref{fig:sub26}, we set a subcritical value of $\tilde{Q}_e$ and see various values of central charge $C$. For the $T-S$ plot we see that for the various values of central charge showing first-order phase transition, at lower entropy values we do not see changes but as the entropy increases we see distinct branches for different values of the $C$. Similarly in the $F-T$ plot for the various values of $C$, we can see there is no change.\\
Similarly, in Figure \ref{fig:sub22}, we fix $\tilde{Q}_e$ at the critical point and vary the central charge $C$, we notice at $\tilde{Q}_e=\tilde{Q}_e^c$, we get a second order phase transition and we see that as $C$ changes for low entropy there is no point able to distinguish the plots but only at higher value of entropy we can. Also at a single value of high entropy, we see that the lesser the value of $C$ the higher the temperature. A similar observation can be seen for Figure \ref{fig:sub24} where we fix for a subcritical value of $\tilde{Q}_e$, we get a first-order phase transition but there is no change in phase as we change the value of $\lambda$. We notice in the plots \ref{fig:sub23} and in \ref{fig:sub24} we see that the plots change the same way when we change the central charge $C$ or the R\'enyi parameter $\lambda$ one at a time, we see that they change the same way showing some kind of relationship which can further be studied.\\
Now when we set both the electric charge $\tilde{Q}_e$ and the magnetic charge $\tilde{Q}_m$ as zero, then the black hole reduces to the Schwarzschild-AdS and the plots change drastically.
At $\tilde{Q}_e=\tilde{Q}_m=0$, we get the value of $T$ \eqref{eq:cftall1} as:-
\begin{equation}
T=\frac{ (\pi  C (3 \lambda  S+4)+3 S (5 \lambda  S+4))}{16 \pi ^{3/2} l \sqrt{S C}}
\end{equation}
Using the equation \eqref{eq:differential}, we get the critical value of the entropy as:-
\begin{equation}
\label{eq:ent001}
S_c= \frac{\sqrt{\pi ^2 C^2 \lambda ^2+88 \pi  C \lambda +16}-\pi C \lambda -4}{30 \lambda }
\end{equation}
Therefore the critical temperature is:-
\begin{equation}
\begin{split}
& T_c=\frac{\left(\pi ^2 \left(-C^2\right) \lambda ^2+\pi  C \lambda  \left(\sqrt{C_1}+72\right)+4 \left(\sqrt{C_1}-4\right)\right)}{8 \sqrt{30 C} \pi ^{3/2} \lambda  l \sqrt{\frac{\sqrt{C_1}-\pi  C \lambda -4}{\lambda }}}\\
\end{split}
\end{equation}
By  the relative parameters $t=T/T_c$ and $s=S/S_c$, we can now write the equation of states as:-
\begin{equation}
\begin{split}
& t=\frac{15  \sqrt{C_2\lambda} \left(\frac{s C_2 \left(s C_2+24\right)}{60 \lambda }+\pi  C \left(\frac{1}{10} s C_2)+4\right)\right)}{\left(-\pi ^2 C^2 \lambda ^2+\pi  C \lambda  \left(\sqrt{C_1}+72\right)+4 \left(\sqrt{C_1}-4\right)\right) \sqrt{\frac{s C_2}{\lambda }}}\\
\end{split}
\end{equation}
Values of $C_1$, $C_2$ are in appendix \ref{appendix}.
\begin{figure}[htp]
    \centering
    \begin{subfigure}[b]{0.4\textwidth}
   \includegraphics[scale=0.7]{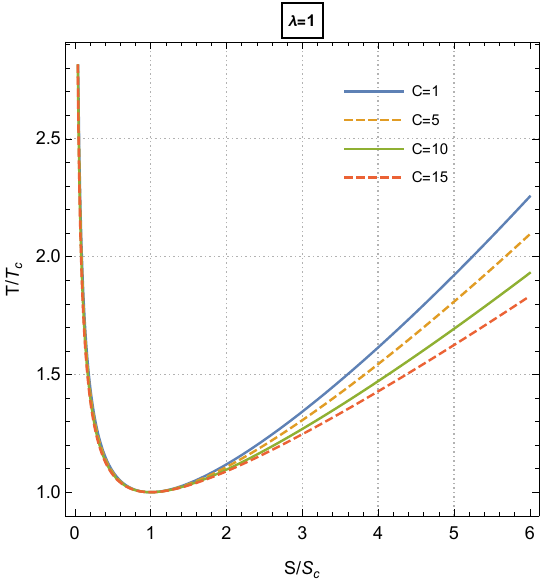}
        \caption{}
        \label{fig:ts001}
    \end{subfigure}
    \hfill
    \begin{subfigure}[b]{0.4\textwidth}
        \includegraphics[scale=0.7]{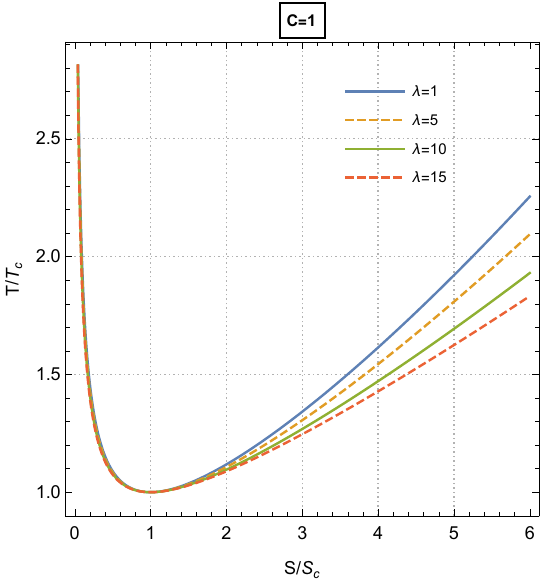}
        \caption{}
        \label{fig:ts002}
    \end{subfigure}
    \caption{$T-S$ plots}
    \label{fig:thirteen}
\end{figure}
For the free energy at $\tilde{Q}_e=\tilde{Q}_m=0$, using \eqref{eq:cftmass1}, \eqref{eq:cftall1} we can write it as:-
\begin{equation}
F=-\frac{\sqrt{S} \sqrt{\frac{l^2}{C}} (\pi  C (\lambda  S-4)+S (9 \lambda  S+4))}{16 \pi ^{3/2} l^2}
\end{equation}
Using equation \eqref{eq:ent001}, we write the critical free energy as:-
\begin{equation}
\begin{split}
& F_c=\frac{\sqrt{\frac{\sqrt{C_1}-\pi C \lambda -4}{\lambda }} \left(\pi ^2 C^2 \lambda ^2-\pi  C \lambda  \left(\sqrt{C_1}-248\right)-4 \left(\sqrt{C_1}-4\right)\right) }{1200 \sqrt{30 C} \pi ^{3/2} \lambda  l}\\
\end{split}
\end{equation}
By writing the free energy in relative parameters $f=F/F_c$ and $s=S/S_c$ we obtain:-
\begin{equation}
\resizebox{1\hsize}{!}{$
f=\frac{\sqrt{\frac{s C_2}{\lambda }} \left(3 s^2 \left(\pi ^2 \left(-C^2\right) \lambda ^2+\pi  C \lambda  \left(\sqrt{C_1}-48\right)+4 \left(\sqrt{C_1}-4\right)\right)-5 s (\pi  C \lambda +4) C_2+600 \pi  C \lambda \right)}{2 \sqrt{\frac{C_2}{\lambda }} \left(\pi ^2 C^2 \lambda ^2-\pi  C \lambda  \left(\sqrt{C_1}-248\right)-4 \left(\sqrt{C_1}-4\right)\right)}$}
\end{equation}

\begin{figure}[htp]
    \centering
    \begin{subfigure}[b]{0.4\textwidth}
   \includegraphics[scale=0.7]{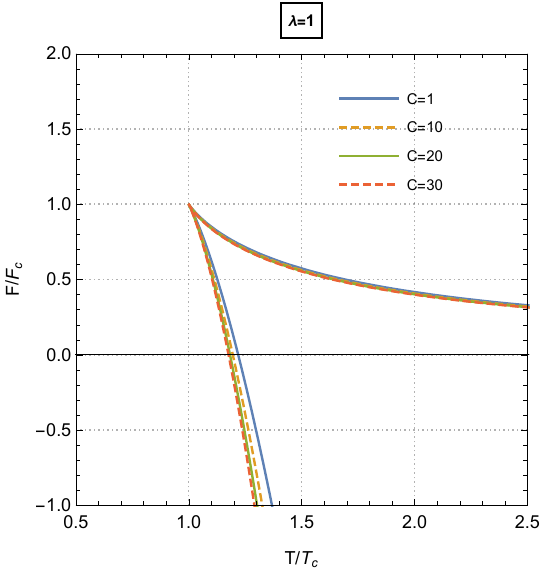}
        \caption{}
        \label{fig:ft001}
    \end{subfigure}
    \hfill
    \begin{subfigure}[b]{0.4\textwidth}
        \includegraphics[scale=0.7]{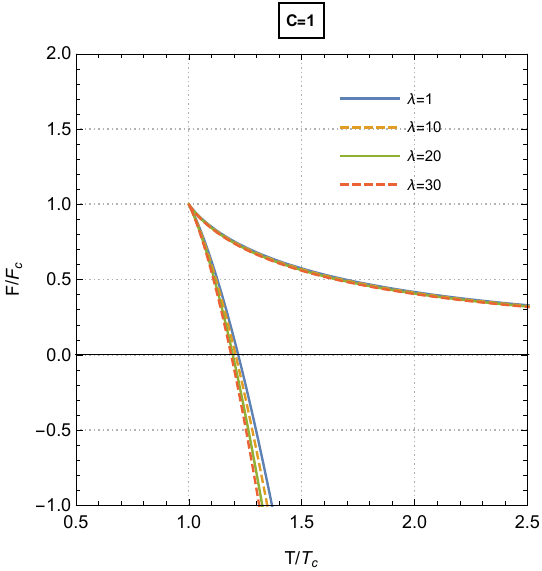}
        \caption{}
        \label{fig:ft002}
    \end{subfigure}
    \caption{$F-T$ plots}
    \label{fig:fourteen}
\end{figure}
The $T-S$ and $F-T$ plots are seen in Figure \ref{fig:thirteen} and the $F-T$ plots are seen in Figure \ref{fig:fourteen}. In Figure \ref{fig:ts001} and \ref{fig:ft001}, we plot for various values of $C$ while keeping $\lambda$ fixed, we see only two branches where for the small entropy we see the unstable branch corresponding small black holes and for the high entropy, we see the stable branch corresponding large black holes. The same structure can be seen in \ref{fig:ts002}, \ref{fig:ft002} when we plot for the various values of $\lambda$ and keep $C$ as fixed. One thing that can be seen in both Figure \ref{fig:ts001} and \ref{fig:ts002} is that at a lower value of entropy, the plots are not distinguishable but at a higher value of entropy we can distinguish for values of $C$ and $\lambda$ where at a particular value of high entropy, higher the value of central charge $C$ or $\lambda$, the lower the temperature $T$. We see that in both the plots in figure \ref{fig:thirteen} and \ref{fig:fourteen} we see that in both the variation of $C$ and $\lambda$ is similar and when $C\rightarrow 0$ is resonates with the plot when $\lambda \rightarrow 0$ which relates to the Bekenstein-Hawking entropy. \\
Also at fixed $\tilde{\Phi}_m$, we can also see the $T-S$ process which can be obtained using equation \eqref{eq:cftall1} and is given as:-
\begin{equation}
T=\frac{ \left(-\pi ^2 (\lambda  S+4) \left(\frac{16 C l^2 S \tilde{\Phi}_m^2}{\pi  (\lambda S-4)^2}+\tilde{Q}_e^2\right)+\pi  C S (3 \lambda  S+4)+3 S^2 (5 \lambda  S+4)\right)}{16 \pi ^{3/2} C^{1/2} l S^{3/2}}
\end{equation}
From equation \eqref{eq:differential}, we get the critical value of $S$ (given in Appendix \ref{appendix}) and $\tilde{\Phi}_m$ as:-

\begin{equation}
\label{eq:pc0011}
\tilde{\Phi}_m^c=\frac{\sqrt{\frac{-45 \lambda  D_1^3-3 (C \pi  \lambda +4) D_1^2+\pi  \left(4 C-\pi  \tilde{Q}_e^2 \lambda \right) D_1-12 \pi ^2 \tilde{Q}_e^2}{l^2 A^{5/2}}}}{4 \sqrt{\pi } \sqrt{\frac{C \left(3 \lambda ^2 D_1^2+24 \lambda  D_1-16\right)}{D_1^{3/2} \left(\lambda  D_1-4\right)^3}}}
\end{equation}
$D_1$ is given in appendix \ref{appendix}.
Using the critical points \eqref{eq:sc0011} and \eqref{eq:pc0011}, we get the critical temperature as:-
\begin{equation}
T_c=\frac{\sqrt{\frac{l^2}{C}} \left(45 \lambda ^3 D_1^5+6 \lambda ^2 (C \pi  \lambda +34) D_1^4-\lambda  \left(\pi ^2 \tilde{Q}_e^2 \lambda ^2-40 C \pi  \lambda +336\right) D_1^3-D_2\right)}{8 l^2 \pi ^{3/2} D_1^{3/2} \left(3 \lambda ^2 D_1^2+24 \lambda  D_1-16\right)}
\end{equation} 

We can write the EOS like:-
\begin{equation}
t=-\frac{D_1^{3/2} \left(3 \lambda ^2 D_1^2+24 \lambda  D_1-16\right)D_4}{2 \left(s A\right)^{3/2} D_5}
\end{equation}
where $D_1$, $D_4$, $D_5$ are given in Appendix \ref{appendix}
where we introduce $t=T/T_c$, $s=S/S_c$ and $\phi_m=\tilde{\Phi}_m/\tilde{\Phi}_m^c$.
\begin{figure}[htp]
    \centering
    \begin{subfigure}[b]{0.4\textwidth}
   \includegraphics[scale=0.7]{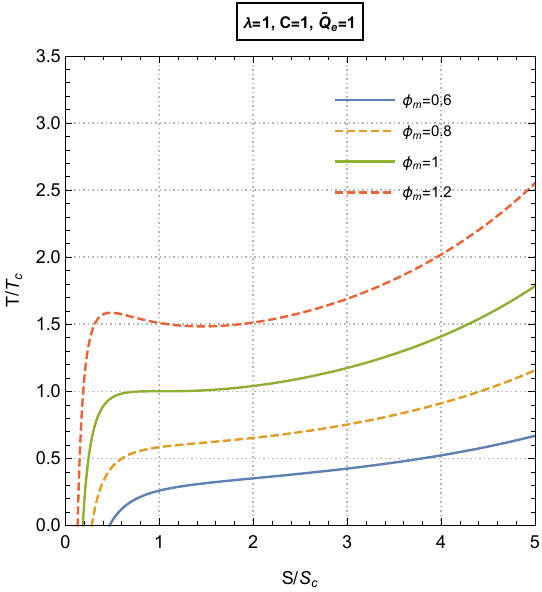}
        \caption{}
        \label{fig:ts011}
    \end{subfigure}
    \hfill
    \begin{subfigure}[b]{0.4\textwidth}
        \includegraphics[scale=0.7]{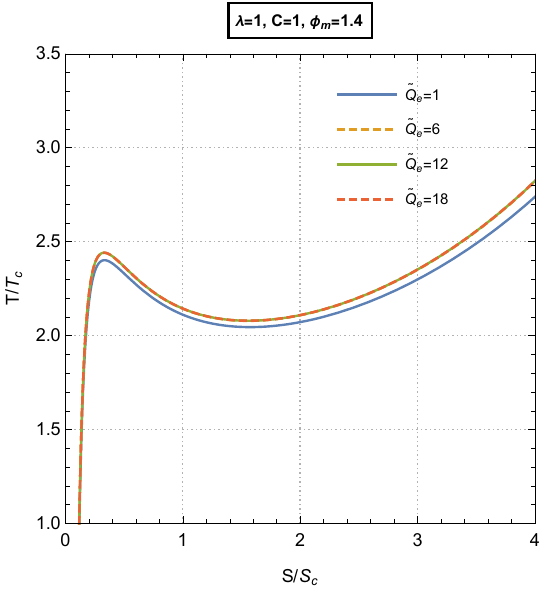}
        \caption{}
        \label{fig:ts012}
    \end{subfigure}
  \begin{subfigure}[b]{0.4\textwidth}
        \includegraphics[scale=0.7]{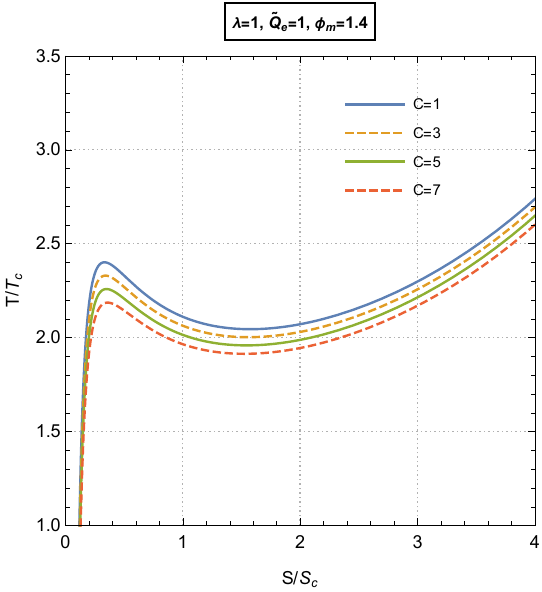}
        \caption{}
        \label{fig:ts013}
    \end{subfigure}
    \hfill
     \begin{subfigure}[b]{0.4\textwidth}
        \includegraphics[scale=0.7]{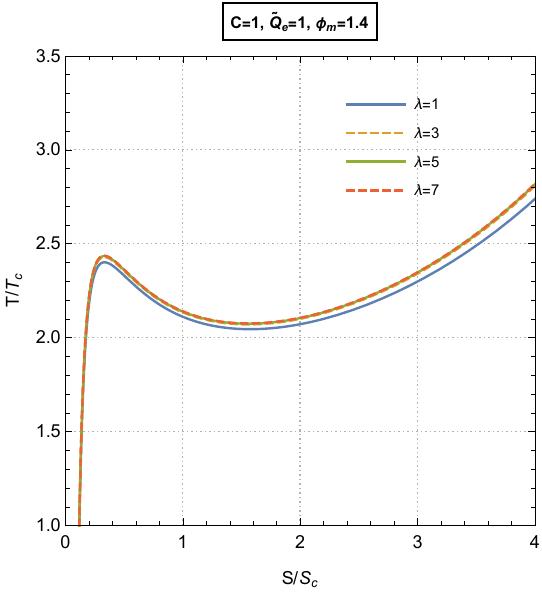}
        \caption{}
        \label{fig:ts014}
    \end{subfigure}
    \caption{$T-S$ plots}
    \label{fig:fifteen}
\end{figure}
Now to plot the $\mu -T$ plot we use equation \eqref{eq:cftall1} and we write $\mu$ in terms of $\tilde{\Phi}_m$ as:-
\begin{equation}
\mu=\frac{\pi  C S \left(\frac{16 l^2 \tilde{\Phi}_m^2}{\lambda  S-4}+\lambda  S+4\right)+\pi ^2 \tilde{Q}_e^2 (\lambda S-4)-S^2 (3 \lambda  S+4)}{16 \pi ^{3/2} C^2 \sqrt{S} \sqrt{\frac{l^2}{C}}}
\end{equation}
Using the critical values from equation \eqref{eq:sc0011} and \eqref{eq:pc0011} we get:-
\begin{equation}
\mu_c=\frac{-27 \lambda ^3 D_1^5+132 \lambda ^2 D_1^4+\lambda  \left(\pi ^2 \tilde{Q}_e^2 \lambda ^2+32 C \pi  \lambda -336\right) D_1^3+D_2}{8 C^2 \sqrt{\frac{l^2}{C}} \pi ^{3/2} \sqrt{D_1} \left(3 \lambda ^2 D_1^2+24 \lambda  D_1-16\right)}
\end{equation}
If $\tilde{m}=\mu/\mu_c$ then we get:-
\begin{equation}
\tilde{m}=\frac{ \left(3 \lambda ^2 D_1^2+24 \lambda  D_1-16\right) D_4}{2 \sqrt{s} D_5}
\end{equation}

\begin{figure}[htp]
    \centering
    \begin{subfigure}[b]{0.4\textwidth}
   \includegraphics[scale=0.7]{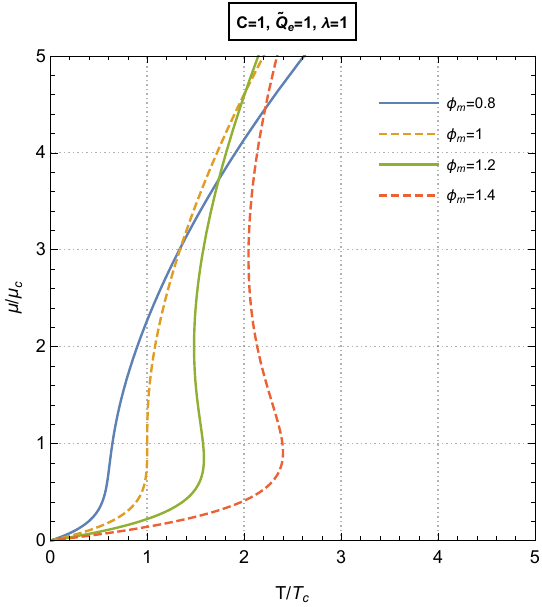}
        \caption{}
        \label{fig:mut01}
    \end{subfigure}
    \hfill
    \begin{subfigure}[b]{0.4\textwidth}
        \includegraphics[scale=0.7]{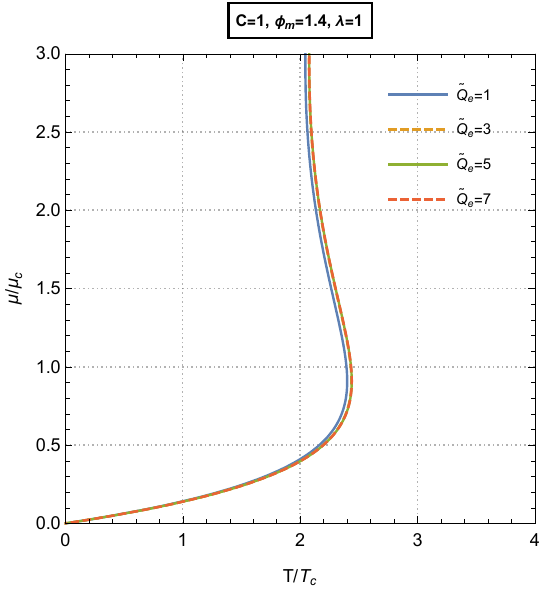}
        \caption{}
        \label{fig:mut02}
    \end{subfigure}
  \begin{subfigure}[b]{0.4\textwidth}
        \includegraphics[scale=0.7]{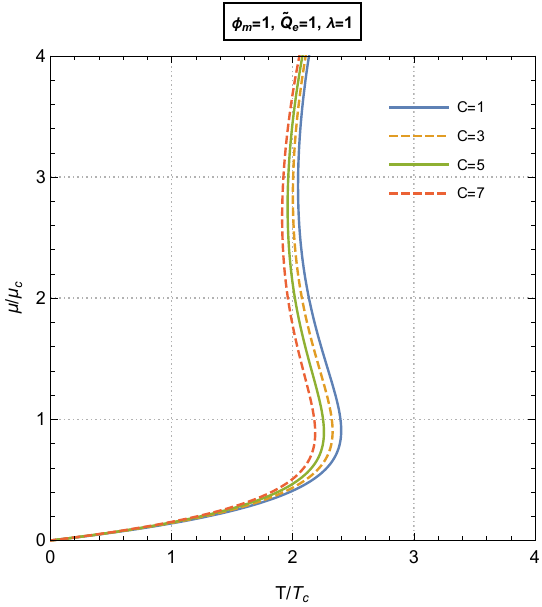}
        \caption{}
        \label{fig:mut03}
    \end{subfigure}
    \hfill
     \begin{subfigure}[b]{0.4\textwidth}
        \includegraphics[scale=0.7]{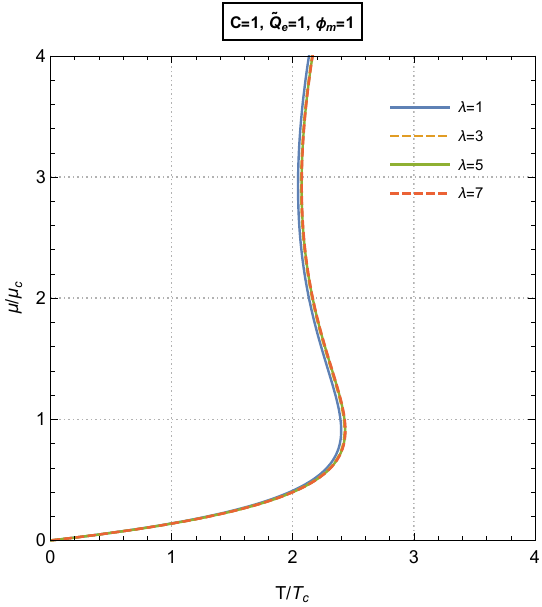}
        \caption{}
        \label{fig:mut04}
    \end{subfigure}
    \caption{$\mu-T$ plots}
    \label{fig:sixteen}
\end{figure}

We plot the $T-S$ and $F-T$ for various values seen in Figure \ref{fig:fifteen}, \ref{fig:sixteen}. In figure (\ref{fig:ts011}), (\ref{fig:mut01}), we see the plot of $T-S$ and $\mu-T$ plot for various fixed values of $\phi_m=\tilde{\Phi}_m/\tilde{\Phi}_m^c$ we see that for sub-critical values of the magnetic potential $\phi_m$, we see a monotonous curve in both the plots signifying two branches one having small entropy (small black holes) and the other branch signifies large entropy (large black hole). At the critical value of $\phi_m$, we see a change of phase of order two and then for the supercritical values of the $\phi_m$, we see a first-order transition similar to the Van der Waals and see three phases namely, small, intermediate and large. In the $\mu-T$ plot for the red plot, at the particular value of $T/T_c=2.2$ which is plotted parametrically with entropy $S$, we see that if a straight line is drawn we see the curve cuts the line at two points and we can see three branches. For the other plots we have chosen a supercritical value of the critical $\phi_m$ and plotted for various values of $\tilde{Q}_e$ in figure (\ref{fig:ts012}), (\ref{fig:mut02}), $C$ in figure (\ref{fig:ts013}), (\ref{fig:mut03}) and $\lambda$ in figure (\ref{fig:ts014}), (\ref{fig:mut04}). We see a first-order phase transition for various values of the three variables and see no transition of phase. If we had plotted for a sub-critical value of the magnetic potential, we would see a monotonous curve showing two phases for different values of the variables. \\
Also for the fixed $\tilde{\Phi}_e$ and $\tilde{\Phi}_m$, we can obtain the value of $T$ from equation \eqref{eq:cftall1} as:-
\begin{equation}
T=\frac{ \left(-\frac{16 \pi  C l^2 (\lambda  S+4) \left(\tilde{\Phi}_e^2+\tilde{\Phi}_m^2\right)}{(\lambda  S-4)^2}+\pi  C (3 \lambda S+4)+3 S (5 \lambda  S+4)\right)}{16 \pi ^{3/2} l \sqrt{C S}}
\end{equation}
From this above equation, we see that there is a single point of extremity for entropy (given in Appendix \ref{appendix}) where the temperature becomes minimum, the points are

\begin{equation}
T_{ex}=\frac{\left(15 \lambda  E_1^2+\left(C \pi  \lambda  \left(3-\frac{16 l^2 \left(\tilde{\Phi}_e^2+\tilde{\Phi}_m^2\right)}{\left(\lambda E_1-4\right)^2}\right)+12\right) E_1+4 C \pi  \left(1-\frac{16 l^2 \left(\tilde{\Phi}_e^2+\tilde{\Phi}_m^2\right)}{\left(\lambda E_1-4\right)^2}\right)\right)}{16 l \pi ^{3/2} \sqrt{E_1 C}}
\end{equation}
where $E_1$ is in Appendix \ref{appendix}.
By introducing some relative variables, we express the EOS as:-
\begin{equation}
\label{eq:tsf}
t=\frac{\left(15 s^2 \lambda  E_1^2+s \left(C \pi  \lambda  \left(3-B\right)+12\right) E_1+4 C \pi  \left(1-B\right)\right)}{\sqrt{s} \left(15 \lambda  E_1^2+\left(C \pi  \lambda  \left(3-D\right)+12\right) E_1+4 C \pi  \left(1-D\right)\right)}
\end{equation}
\begin{equation}
B=\frac{16 l^2 \left(\tilde{\Phi}_e^2+\tilde{\Phi}_m^2\right)}{\left(s \lambda  A-4\right)^2}, \quad D=\frac{16 l^2 \left(\tilde{\Phi}_e^2+\tilde{\Phi}_m^2\right)}{\left(\lambda  A-4\right)^2}
\end{equation}
where $s=S/S_{ex}$ and $t=T/T_{ex}$. Similarly, the chemical potential reaches the highest value at the entropy $S=S_{ex}$. The $\mu$ is written as:-
\begin{equation}
\mu=\frac{\sqrt{S} \left(\pi  C \left(16 l^2 \left(\tilde{\Phi}_e^2+\tilde{\Phi}_m^2\right)+\lambda ^2 S^2-16\right)+S \left(-3 \lambda ^2 S^2+8 \lambda  S+16\right)\right)}{16 \pi ^{3/2} C^2 \sqrt{\frac{l^2}{C}} (\lambda S-4)}
\end{equation} 
And its critical value is:-
\begin{equation}
\mu_{ex}=\frac{\sqrt{E_1} \left(C \pi  \left(16 \left(\tilde{\Phi}_e^2+\tilde{\Phi}_m^2\right) l^2+\lambda ^2 E_1^2-16\right)+E_1 \left(-3 \lambda ^2 E_1^2+8 \lambda  E_1+16\right)\right)}{16 C^2 \sqrt{\frac{l^2}{C}} \pi ^{3/2} \left(\lambda  E_1-4\right)}
\end{equation}
Using the relative variable $\tilde{m}=\mu/\mu_{ex}$ we get:-
\begin{equation}
\label{eq:msf}
\tilde{m}=-\frac{\sqrt{s} \left(\lambda  E_1-4\right) \left(3 s^3 \lambda ^2 E_1^3-s^2 \lambda  (C \pi  \lambda +8) E_1^2-16 s E_1-16 C \pi  \left(l^2 \left(\tilde{\Phi}_e^2+\tilde{\Phi}_m^2\right)-1\right)\right)}{ \left(s \lambda  E_1-4\right) \left(-3 \lambda ^2E_1^3+\lambda  (C \pi  \lambda +8)E_1^2+16 E_1+16 C \pi  \left(l^2 \left(\tilde{\Phi}_e^2+\tilde{\Phi}_m^2\right)-1\right)\right)}
\end{equation}

\begin{figure}[htp]
    \centering
    \begin{subfigure}[b]{0.4\textwidth}
   \includegraphics[scale=0.7]{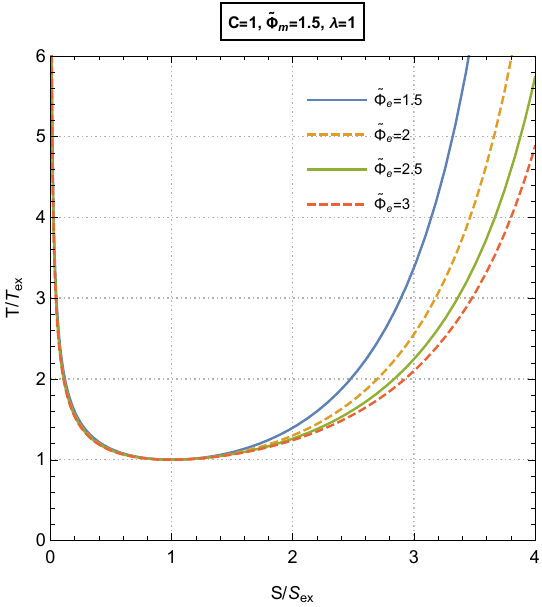}
        \caption{}
        \label{fig:ts021}
    \end{subfigure}
    \hfill
    \begin{subfigure}[b]{0.4\textwidth}
        \includegraphics[scale=0.7]{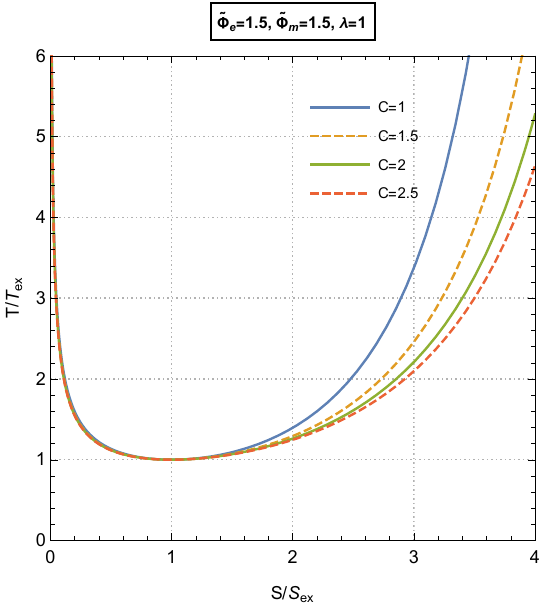}
        \caption{}
        \label{fig:ts022}
    \end{subfigure}
  \begin{subfigure}[b]{0.4\textwidth}
        \includegraphics[scale=0.7]{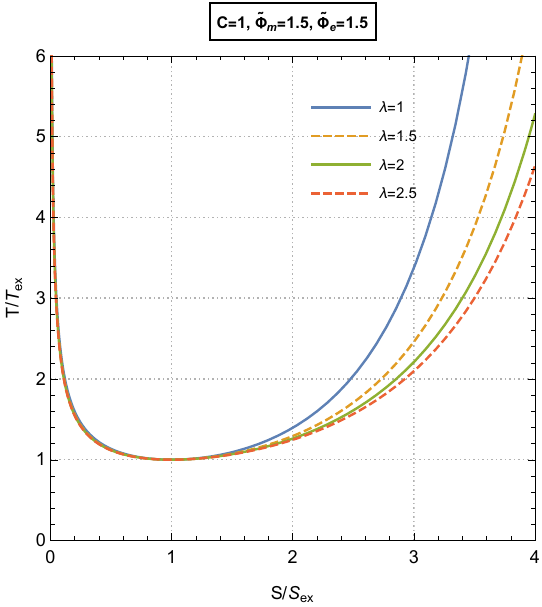}
        \caption{}
        \label{fig:ts023}
    \end{subfigure}
    
    \caption{$T-S$ plots}
    \label{fig:seventeen}
\end{figure}

\begin{figure}[htp]
    \centering
    \begin{subfigure}[b]{0.4\textwidth}
   \includegraphics[scale=0.7]{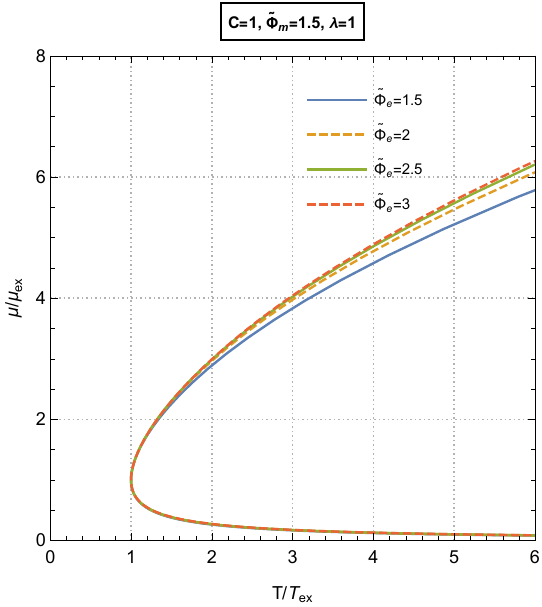}
        \caption{}
        \label{fig:mut011}
    \end{subfigure}
    \hfill
    \begin{subfigure}[b]{0.4\textwidth}
        \includegraphics[scale=0.7]{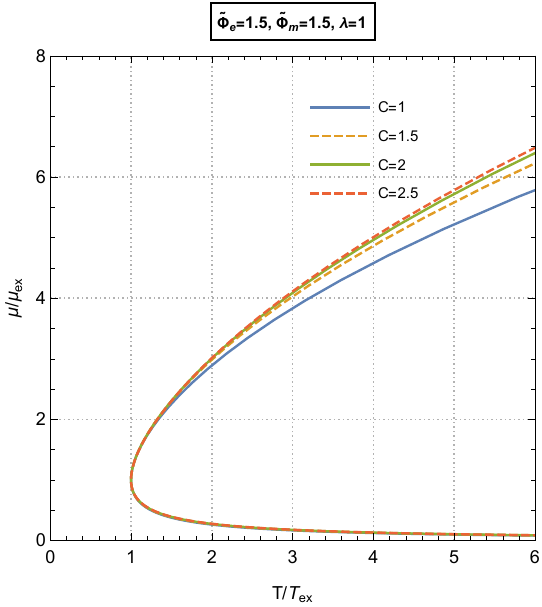}
        \caption{}
        \label{fig:mut012}
    \end{subfigure}
  \begin{subfigure}[b]{0.4\textwidth}
        \includegraphics[scale=0.7]{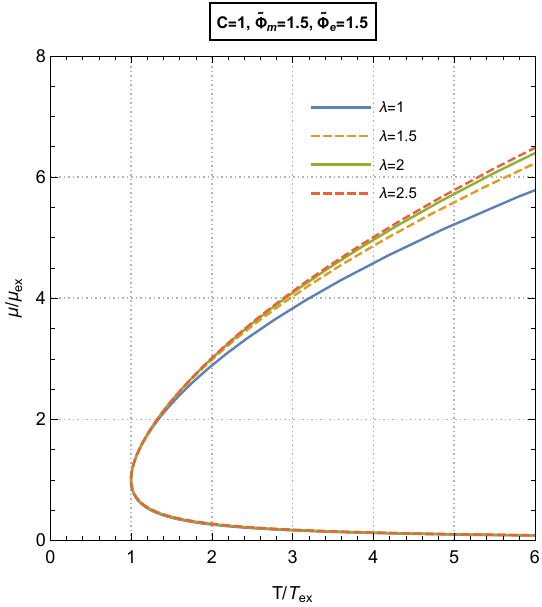}
        \caption{}
        \label{fig:mut013}
    \end{subfigure}
    
    \caption{$\mu-T$ plots}
    \label{fig:eighteen}
\end{figure}
\begin{figure}[htp]
    \centering
    \begin{subfigure}[b]{0.4\textwidth}
   \includegraphics[scale=0.8]{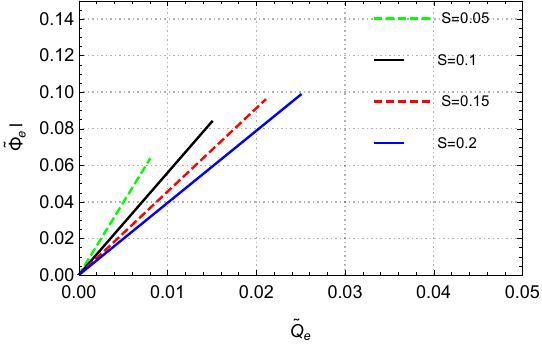}
        \caption{$\tilde{\Phi}_e-\tilde{Q}_e$ plots}
        \label{fig:pq1}
    \end{subfigure}
    \hfill
    \begin{subfigure}[b]{0.4\textwidth}
        \includegraphics[scale=0.8]{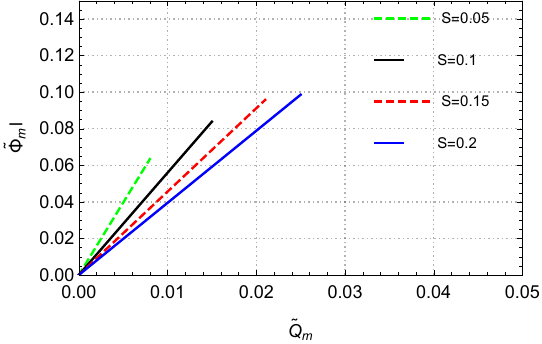}
        \caption{$\tilde{\Phi}_m-\tilde{Q}_m$ plots}
        \label{fig:pq2}
    \end{subfigure}

    \caption{}
    \label{fig:nineteen}
\end{figure}

We can see that both the equation of state \eqref{eq:tsf} and the chemical potential equation \eqref{eq:msf} have a dependency with $\tilde{\Phi}_e$, $\tilde{\Phi}_m$ as well as the entropy $S$ and the central charge $C$. We see the $T-S$ plots in Figure \ref{fig:seventeen} and the $\mu-T$ plots in Figure \ref{fig:eighteen}. In the plot (\ref{fig:ts021}) and (\ref{fig:mut011}), we plot for various values of electric potential $\tilde{\Phi}_e$. We see only two branches for both the plots and we do not see much change for various values of the electric potential. In the $T-S$ plot we do not see much changes for various values in low entropy but we can distinguish it in higher entropy $S$. In Figure \ref{fig:mut011}, we also see two branches in the $\mu-T$ plot. In other plots also namely with various values of $C$ in Figure (\ref{fig:ts022}), (\ref{fig:mut012}) and also for various values of $\lambda$ in Figure (\ref{fig:ts023}), (\ref{fig:mut013}). One thing different is that in the Bekenstein Hawking entropy construct, the plots for the $\tilde{Q}_e= \tilde{Q}_m=0$ condition is the same for the canonical ensemble namely when replacing the dual charges with its potentials but in the R\'enyi entropy construct through the plot when $\tilde{Q}_e=\tilde{Q}_m=0$ and during an isovoltage process is not same but their features are similar that both show two phases with small entropy which corresponds for unstable black hole and for large entropy which corresponds for a stable black hole. \\
We can also check for the thermodynamical process for $\tilde{\Phi}_e-\tilde{Q}_e$ and $\tilde{\Phi}_m-\tilde{Q}_m$ plots. But if we look closely at equation \eqref{eq:cftall1}, we notice its direct proportionality of the potentials to the charges so the thermodynamical plots do not show any points of inflection or points of extremity. We can see the plots in figure \ref{fig:nineteen}.\\
For the $\mu-C$ process, we start from equation \eqref{eq:cftall1} and write the equation of $\mu$. When we graph the $\mu-C$ plot we see the point of extremity for fixed values. Therefore the points of extremity are:-
\begin{equation}
\begin{split}
& C_{max}=\frac{3 S^2 (3 \lambda  S+4)-3 \pi ^2 \left(\tilde{Q}_e^2+\tilde{Q}_m^2\right) (\lambda  S-4)}{\pi  S (\lambda  S+4)}\\
& \mu_{max}=\frac{\sqrt{S} (\lambda  S+4) \sqrt{\frac{l^2 S (\lambda  S+4)}{S^2 (3 \lambda  S+4)-\pi ^2 \left(\tilde{Q}_e^2+\tilde{Q}_m^2\right) (\lambda S-4)}}}{24 \sqrt{3} l^2}.
\end{split}
\end{equation}
We give dimension parameters $c=C/C_{max}$ and $m=\mu /\mu_{max}$. By using the parameters, the $\mu$ equation becomes:-
\begin{equation}
m=\frac{3 \mathit{c}-1}{2 \mathit{c}^2 \sqrt{\frac{1}{\mathit{c}}}}
\end{equation}
\begin{figure}[h]
\centering
\includegraphics[scale=0.7]{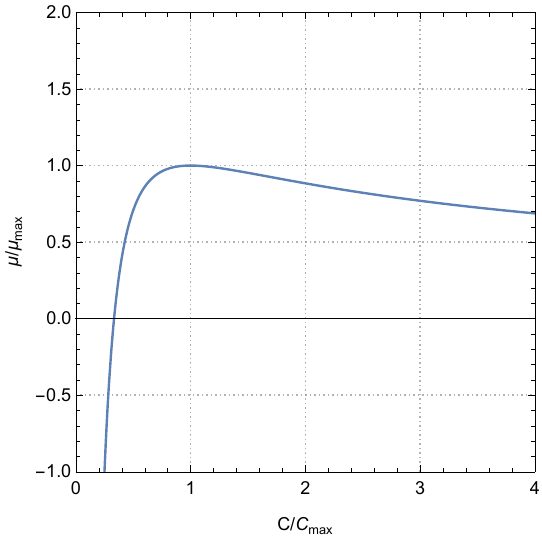}
\caption{$\mu-C$ process for fixed values of ($\tilde{Q}_e, \tilde{Q}_m,S$)}
\label{fig:twenty}
\end{figure}
We can see from Figure \ref{fig:twenty} for and fixed values, we get the same $\mu-C$ plot which is the same as obtained in the previous section using the Bekenstein-Hawking entropy signaling some kind of universality of the $\mu-C$ not between different types of black holes namely Reissner–Nordström, Kerr, Kerr Sen, and other black holes but also with different entropy constructs.
\section{Conclusions}
\label{sec:five}
In this paper, we try to study the Restricted Phase space thermodynamics of AdS black holes using two different entropy constructs, namely the Bekestein-Hawking entropy model and the R\'enyi entropy model, and examine if by changing the entropy models do we see any change in the thermodynamics using the RPST formalism. In Section \ref{sec:one}, we give a bit of introduction for using the RPST formalism here since there is a problem with the Extended Phase Space Thermodynamics and then we discuss briefly how the dyonic system due to the addition of an extra parameter namely the magnetic charge magnifies the platform for the study of phase structures by increasing the number of ensembles and also introducing what we call the mixed ensembles. This introduction makes the phase structure more rich and vivid. Also, we discuss how the entropy of a black hole is non-extensive and many scientists have tried to model out different constructs for the entropy since the thermodynamics and structures of the black hole depend largely on the entropy so maybe one could see a difference in using different entropy models and our main idea or motivation is to study and compare the thermodynamics of the AdS black hole using two entropy models.\\
In Section \ref{sec:two} we deduce the preliminary calculations for the study of bulk thermodynamics and then state the rescaled electric and magnetic charge and potential respectively. We also introduce the central charge $C$ which is crucial for the study of black holes in the boundary which incorporates the change in Newton's constant while keeping the AdS length fixed- the ground basis of the RPST formalism.\\
In section \ref{sec:three}, we study the thermodynamics using the Bekenstein-Hawking entropy calculate the various thermodynamic values required in the RPST formalism, and also check its homogeneity. The mass equation $M$ is of order one and for the rest variables it is all zero. Next, we study the various thermodynamic processes. In Figure \ref{fig:one} we see different phase transitions namely the Van der Waals and then the non-equilibrium phase transition. We can see that for the sub-critical value of $q_e$ we get a monotonous graph stating two branches corresponding to the small and large black hole, At the critical value of $q_e$ we see a change of phase of order two and then for the supercritical values of $q_e$, we see three branches in both the $T-S$ and $F-T$ plots and then as the value of $q_e$ increases, there is again a phase transition where we see only two branches. We do not see the addition of the Hawking-Page transition in Figure \ref{fig:two} because we have turned one charge to zero, stating the effect of dual charge on the phase transition. But the order of transition interchanges as we see in Figure \ref{fig:one} the first order phase transition occurs for supercritical values and for Figure \ref{fig:two} the first order phase transition occurs for subcritical values. In Figure \ref{fig:three}, we see only two phases and it has already been seen in the paper studied before \cite{rps}. In Figure \ref{fig:four}, we replace the electric charge $\tilde{Q}_e$ with the electric potential $\tilde{\Phi}_e$, in doing so we achieve a new mixed ensemble which would not have been possible in a single charge black hole. We see that there is a phase transition when one goes from the subcritical to the supercritical values of the electric potential $\tilde{\phi}_e$. We see in both the plots of $T-S$ and $\mu-T$ that there is a phase transition at the critical point where the number of branches changes from two to three signaling an intermediate branch creation. In the grand canonical ensemble the plots in Figure \ref{fig:six} interestingly seem to be the same as the Schwarzschild black hole case linking to some kind of similar of this grand canonical ensemble to the Schwarzschild black hole. Lastly, we see the $\tilde{\Phi}_e-\tilde{Q}_e$, $\tilde{\Phi}_m-\tilde{Q}_m$ process where it is not much of interest since the potentials are in direct proportion to the charge hence they show no point of inflection or criticality. Finally in Figure \ref{fig:eight}, we see the $\mu-C$ process which is quite similar to many papers studied in this RPST formalism having charge, spin, both charge and spin and now dual charge summarising that there might be some underlining universality in this process.\\
In section \ref{sec:four} we study thermodynamics using the R\'enyi entropy model and calculate the various thermodynamic values required to study different processes and also check homogeneity. In Figure \ref{fig:nine} and Figure \ref{fig:ten}, we see the $T-S$ and $F-T$ plots for various variables. Comparing with Figure \ref{fig:one}, we see both Van der Walls phase transitions while going from subcritical to supercritical values of variables taken in the plot, and at a particular supercritical value of the variable we see a non-equilibrium and Hawking type phase transition in both the $T-S$ and $F-T$ plot respectively across the entropy models. In Figure \ref{fig:eleven} and \ref{fig:twelve}, we see the devoid of the Hawking-page transition when one charge $\tilde{Q}_m=0$, while the Van der Waals type phase transition and the phenomenon is similar for both the entropy constructs. when both the charges are set to zero in Figure \ref{fig:thirteen}, we see only two branches which is similar to the plot seen in Figure \ref{fig:three}. Till here the phase transition is quite similar across the entropy models. Now for the mixed ensemble by fixing $\tilde{\Phi}_m$, we have plotted the $T-S$ and the $\mu-T$ plot in Figure \ref{fig:fifteen} and \ref{fig:sixteen}. We see that the $T-S$ plots are quite similar to the Bekenstein entropy seen in Figure \ref{fig:four}, but there is s slight dissimilarity with the $\mu-T$ plot. We do see Phase transition from two phases to three phases in both the plots but if in \ref{fig:four} we see that the plots start from a finite value of $\mu$ cuts the x-axis at a finite value of $T$. Thus here the entropy peaks at a certain value and then diminishes to zero at a particular temperature but in Figure \ref{fig:sixteen}, we see that as the $\mu$ increases, the value of temperature first increases then decreases and then it keeps on increasing for the supercritical values of $\tilde{\Phi}_m$. We can infer from here that there is a dissimilarity in the phase diagram in this mixed ensemble over the change of the entropy models which would not be visible had one have not taken a dyonic black hole system. In figure \ref{fig: seventeen}, the grand canonical ensemble of the $T-S$ plot is similar to figure \ref{fig:thirteen} but again the $\mu-T$ is dissimilar with the plot where both charges are set to zero which was similar with the Bekensetein entropy setup. In Figure \ref{fig:nineteen}, we do not see any inflection and in Figure \ref{fig:twenty}, the $\mu-C$ process is similar even with the change in entropy models.\\
The study of a dyonic black hole in this RPST formalism showed amazing novel features and also resembled the RN-AdS, Kerr AdS, and Kerr-Sen AdS as studied before. We summarise the paper in the following points-\\
{\bf 1.}The study of the dyonic black holes in the RPST formalism relates to the standard description of Extensive Phase Space Thermodynamics. Here the mass is a homogeneous function of all the variables and is of the order one whereas rest of the variables is of the order zero and it is the internal energy. Since there is no pressure-volume term, we cannot define the mass as enthalpy.\\
{\bf 2.}For all the $T-S$ processes for the fixed electric charge, we get a first-order phase transition but due to the addition of the magnetic charge we see a non-equilibrium transition for the unstable small black hole to the stable large black hole and the Hawking-Page phase transition is seen in $F-T$ plot over the different models of entropy namely the Bekenstein and the R\'enyi which was not seen for a single charged black hole. \\
{\bf 3.} Due to magnetic charge $\tilde{Q}_m$ addition, we got to study a mixed ensemble $(\tilde{\Phi}_e, \tilde{Q}_m)$ which will not be seen in a single charge and we obtained Van der Waals type transition of phase in this ensemble seen in Figure \ref{fig:four}, \ref{fig:fifteen}, \ref{fig:sixteen}.\\
{\bf 4}. In this mixed ensemble only we see the change of plots namely the $F-T$ and $\mu-T$ plots as the models of entropy change even though the nature of the transition remains the same. In the Bekenstein-Hawking entropy, there is similarity in the $F-T$ and $\mu-T$ plots but there is a difference in the R\'enyi entropy model between the two plots.\\
{\bf 5.}In the $\tilde{\Phi}_e-\tilde{Q}_e$ and also for the $\tilde{\Phi}_m-\tilde{Q}_m$ processes at a fixed entropy $S$ is quite trivial as there is no criticality of phase transitions.\\
{\bf 6.} One thing we noticed is that when $\lambda \rightarrow 0$ in the R\'enyi entropy model, it comes back to the Bekenstein-Hawking entropy and its changes in value cause its diversion from the normal plot obtained using the Bekenstein-Hawking entropy plots. We see that the variation of central charge $C$ also mimics the same kind of deviations seen in numerous figures using the R\'enyi entropy signaling some kind of similarity.\\
{\bf 7.} The $\mu-C$ plot remains the same even in different black hole systems seen in different papers and also in different entropy models, sparking a universality in the process.

\section{Appendix}
\label{appendix}
Since most of the calculations we done using the software Mathematica, some of the solutions were not of the usual format and were quite repetitive and to keep the paper concise, we wrote down the expressions here.
\begin{equation}
A_1=A_2-\lambda  (C \pi  \lambda +544)+\frac{\lambda ^2 (C \pi  \lambda -536)^2}{A_2}
\end{equation}
\begin{equation}
\small
\begin{split}
&A_2=\sqrt[3]{-C^3 \pi ^3 \lambda ^6+1608 C^2 \pi ^2 \lambda ^5+1082112 C \pi  \lambda ^4-153938944 \lambda ^3+360 \sqrt{6} A_3}\\
&A_3=\sqrt{-\lambda ^6 \left(5 C^4 \pi ^4 \lambda ^4-8832 C^3 \pi ^3 \lambda ^3+722976 C^2 \pi ^2 \lambda ^2+87079424 C \pi  \lambda +20477952\right)}
\end{split}
\end{equation}
\begin{equation}
A_4=\frac{-\pi ^2 \left(\frac{A}{90 \lambda }+12\right) \tilde{Q}_m^2-\frac{\left(A_1\right)^2 \left(\frac{A_1}{6 \lambda }+4\right)}{2700 \lambda ^4}+\frac{C \pi  \left(A_1\right) \left(4-\frac{A_1}{30 \lambda }\right)}{90 \lambda ^2}}{\pi ^2 \left(\frac{A_1}{90 \lambda }+12\right)}
\end{equation}
\begin{equation}
A_5=\frac{3 \left(\frac{A_1}{90 \lambda }-12\right) \left(C \pi  \lambda  \left(+120 \lambda - A_1 \right)-A_1 \left(\frac{A_1}{6 \lambda }+4\right)\right)}{\frac{A_1}{90 \lambda }+12}
\end{equation}

\begin{equation}
B_1=-\pi ^3 C^3 \lambda ^6+1608 \pi ^2 C^2 \lambda ^5+360 \sqrt{6} \sqrt{B_2}+1082112 \pi  C \lambda ^4-153938944 \lambda ^3
\end{equation}
\begin{equation}
 B_2=-\lambda ^6 \left(5 \pi ^4 C^4 \lambda ^4-8832 \pi ^3 C^3 \lambda ^3+722976 \pi ^2 C^2 \lambda ^2+87079424 \pi  C \lambda +20477952\right)
\end{equation}
 \begin{equation}
 B_3=\sqrt[3]{B_1}-\lambda  (C \pi  \lambda +544)+\frac{\lambda ^2 (C \pi  \lambda -536)^2}{\sqrt[3]{B_1}}
 \end{equation}
 
 \begin{equation}
 B_4=30 \lambda ^2 \left(\frac{B_3}{90 \lambda }+4\right) \left(C \pi  \left(4-\frac{B_3}{30 \lambda }\right)-\frac{\left(B_3\right) \left(\frac{B_3}{6 \lambda }+4\right)}{30 \lambda ^2}\right)
 \end{equation}
 
 \begin{equation}
 C_1=\pi ^2 C^2 \lambda ^2+88 \pi  C \lambda +16
\end{equation}  
 \begin{equation}
 C_2=\sqrt{C_1}-\pi c \lambda -4
 \end{equation}

\begin{equation}
\begin{split}
&D_1= A=\text{Root}\left[135 \lambda ^4 \text{$\#$1}^6+\left(6 C \pi  \lambda ^4+1464 \lambda ^3\right) \text{$\#$1}^5+\left(\pi ^2 \tilde{Q}_e^2 \lambda ^4+80 C \pi  \lambda ^3-2304 \lambda ^2\right) \text{$\#$1}^4\right.\\
&\left. +\left(24 \pi ^2 \tilde{Q}_e^2 \lambda ^3-160 C \pi  \lambda ^2+1664 \lambda \right) \text{$\#$1}^3+\left(256 \pi ^2 \tilde{Q}_e^2 \lambda ^2+256\right) \text{$\#$1}^2+\right.\\
&\left.640 \pi ^2 \tilde{Q}_e^2 \lambda  \text{$\#$1}-256 \pi ^2 \tilde{Q}_e^2\&,2\right]
\end{split}
\end{equation} 
 
 \begin{equation}
D_2=B=12 \left(\pi ^2 \tilde{Q}_e^2 \lambda ^2+16\right) A^2-48 \pi ^2 \tilde{Q}_e^2 \lambda A-64 \pi ^2 \tilde{Q}_e^2
 \end{equation}
 
 \begin{equation}
 D_3=\frac{s \phi_m^2 \left(\lambda   D_1-4\right)^3 \left(45 \lambda  D_1^3+3 (C \pi  \lambda +4)  D_1^2+\pi  \left(\pi  \tilde{Q}_e^2 \lambda -4 C\right)  D_1+12 \pi ^2 \tilde{Q}_e^2\right)}{\pi ^2 \left(s \lambda   D_1-4\right)^2 \left(3 \lambda ^2  D_1^2+24 \lambda  A-16\right)}
 \end{equation}
 
 \begin{equation}
 \begin{split}
& D_4= \left(3 s^2 \left(5 s \lambda  D_1+4\right) D_1^2+C \pi  s \left(3 s \lambda  D_1+4\right) D_1-\right.
 \left.\pi ^2 \left(s \lambda   D_1+4\right) \left(\tilde{Q}_e^2- D_3\right)\right)\\
& D_5=-45 \lambda ^3  D_1^5-6 \lambda ^2 (C \pi  \lambda +34)  D_1^4+\lambda  \left(\pi ^2 \tilde{Q}_e^2 \lambda ^2-40 C \pi  \lambda +336\right)  D_1^3\\
&+12 \left(\pi ^2 \tilde{Q}_e^2 \lambda ^2+16\right) D_1^2+48 \pi ^2 \tilde{Q}_e^2 \lambda  D_1+64 \pi ^2 \tilde{Q}_e^2
\end{split}
 \end{equation}
 
 \begin{equation}
\begin{split}
&E_1= \text{Root}\left[45 \lambda ^4 \text{$\#$1}^5+\left(3 C \pi  \lambda ^4-528 \lambda ^3\right) \text{$\#$1}^4+\left(2016 \lambda ^2-40 C \pi  \lambda ^3\right) \text{$\#$1}^3\right.\\
&\left.+\left(48 C l^2 \pi  \tilde{\Phi}_e^2 \lambda ^2+48 C l^2 \pi  \tilde{\Phi}_m^2 \lambda ^2+192 C \pi  \lambda ^2-2304 \lambda \right) \text{$\#$1}^2\right.\\
&\left.+\left(384 C l^2 \pi  \lambda  \tilde{\Phi}_e^2+384 C l^2 \pi  \lambda  \tilde{\Phi}_m^2-384 C \pi  \lambda -768\right) \text{$\#$1}-\right.\\
&\left.256 c l^2 \pi  \tilde{\Phi}_e^2-256 c l^2 \pi  \tilde{\Phi}_m^2+256 C \pi \&,1\right]
\end{split}
\end{equation}
 
\begin{equation}
\label{eq:sc0011}
\begin{split}
& S_c= \text{Root}\left[\text{$\#$1}^5 \left(6 \pi  C \lambda ^4+1464 \lambda ^3\right)+\text{$\#$1}^4 \left(80 \pi  C \lambda ^3-2304 \lambda ^2+\pi ^2 \lambda ^4 \tilde{Q}_e^2\right)\right.\\
& \left.+\text{$\#$1}^3 \left(-160 \pi  C \lambda ^2+1664 \lambda +24 \pi ^2 \lambda ^3 \tilde{Q}_e^2\right)\right.\\
&\left.+135 \text{$\#$1}^6 \lambda ^4+\text{$\#$1}^2 \left(256 \pi ^2 \lambda ^2 \tilde{Q}_e^2+256\right)+640 \pi ^2 \text{$\#$1} \lambda  \tilde{Q}_e^2-256 \pi ^2 \tilde{Q}_e^2\&,2\right]
\end{split}
\end{equation} 
 
\begin{equation}
\begin{split}
& S_{ex}=\text{Root}\left[\text{$\#$1}^4 \left(3 \pi  C \lambda ^4-528 \lambda ^3\right)+\text{$\#$1}^3 \left(2016 \lambda ^2-40 \pi  C \lambda ^3\right)\right. \\
& +\left.\text{$\#$1}^2 \left(192 \pi  C \lambda ^2+48 \pi  C \lambda ^2 l^2 \tilde{\Phi}_e^2+48 \pi  C \lambda ^2 l^2 \tilde{\Phi}_m^2-2304 \lambda \right)+45 \text{$\#$1}^5 \lambda ^4\right.\\
&\left.+\text{$\#$1} \left(-384 \pi  C \lambda +384 \pi  C \lambda  l^2 \tilde{\Phi}_e^2+384 \pi  C \lambda  l^2 \tilde{\Phi}_m^2-768\right)\right.\\
& \left.-256 \pi  C l^2 \tilde{\Phi}_e^2-256 \pi  C l^2 \tilde{\Phi}_m^2+256 \pi  C\&,1\right]
\end{split}
\end{equation}

\end{document}